\documentclass[fleqn,usenatbib]{mnras}

\usepackage[T1]{fontenc}

\DeclareRobustCommand{\VAN}[3]{#2}
\let\VANthebibliography\thebibliography
\def\thebibliography{\DeclareRobustCommand{\VAN}[3]{##3}\VANthebibliography}

\usepackage{graphicx}	
\usepackage{amsmath}	
\usepackage{amssymb}	
\usepackage{color}
\usepackage{hyperref}
\usepackage{siunitx}
\usepackage{float}
\usepackage[dvipsnames]{xcolor}

\usepackage{newtxtext,newtxmath}

\newcommand{\mtwo}{$M_{\text{200m}}$}
\newcommand{\mstr}{$M_\star$}

\newcommand{\rtwo}{$R_{\text{200m}}$}
\newcommand{\msun}{M$_{\odot}$}

\newcommand{\lcdm}{$\Lambda$CDM}

\newcommand{\vmax}{$V_\text{max}$}

\newcommand{\rockstar}{\texttt{ROCKSTAR}}

\newcommand{\tquench}{$t_\text{quench}$}

\newcommand{\e}[1]{$\times 10^ {#1}$}

\title[Environmental effects of LMC-mass hosts]{The effects of LMC-mass environments on their dwarf satellite galaxies in the FIRE simulations}

\author[E.D. Jahn et al.]{
Ethan D. Jahn,$^{1}$\thanks{E-mail: ejahn003@ucr.edu}
Laura V. Sales,$^{1}$
Andrew Wetzel,$^{2}$
Jenna Samuel,$^{2}$
Kareem El-Badry,$^{3}$\newauthor
Michael Boylan-Kolchin,$^{4}$
and James S. Bullock$^{5}$
\\
$^{1}$Department of Physics and Astronomy, University of California, Riverside, CA 92507, USA\\
$^{2}$Department of Physics and Astronomy, University of California, Davis, CA 95616, USA\\
$^{3}$Department of Astronomy and Theoretical Astrophysics Center, University of California Berkeley, Berkeley, CA 94720, USA\\
$^{4}$Department of Astronomy, The University of Texas at Austin, 2515 Speedway Blvd, Stop C1400, Austin, TX 78712, USA\\
$^{5}$Department of Physics and Astronomy, University of California, Irvine, CA 92697, USA
}

\date{Accepted XXX. Received YYY; in original form ZZZ}

\pubyear{2015}

\begin{document}
\label{firstpage}
\pagerange{\pageref{firstpage}--\pageref{lastpage}}
\maketitle

\begin{abstract}
Characterizing the predicted environments of dwarf galaxies like the Large Magellanic Cloud (LMC) is becoming increasingly important as next generation surveys push sensitivity limits into this low-mass regime at cosmological distances. We study the environmental effects of LMC-mass halos ($M_{200m} \sim 10^{11}$ M$_\odot$) on their populations of satellites ($M_\star \geq 10^4$ M$_\odot$) using a suite of zoom-in simulations from the Feedback In Realistic Environments (FIRE) project. Our simulations predict significant hot coronas with $T\sim10^5$ K and $M_\text{gas}\sim10^{9.5}$ M$_\odot$. We identify signatures of environmental  quenching in dwarf satellite galaxies, particularly for satellites with intermediate mass ($M_\star = 10^{6-7}$ M$_\odot$). The gas content of such objects indicates ram-pressure as the likely quenching mechanism, sometimes aided by star formation feedback. Satellites of LMC-mass hosts replicate the stellar mass dependence of the quiescent fraction found in satellites of MW mass hosts (i.e. that the quiescent fraction increases as stellar mass decreases). Satellites of LMC-mass hosts have a wider variety of quenching times when compared to the strongly bi-modal distribution of quenching times of nearby centrals. Finally, we identify significant tidal stellar structures around four of our six LMC-analogs, suggesting that stellar streams may be common. These tidal features originated from satellites on close orbits, extend to $\sim$80 kpc from the central galaxy, and contain $\sim10^{6-7}$ M$_\odot$~of stars.
\end{abstract}

\begin{keywords}
galaxies: dwarf -- galaxies: formation -- galaxies: evolution -- galaxies: star formation -- Local Group
\end{keywords}



\section{Introduction}

Dwarf galaxies, due to their shallow potentials and low baryon fractions, represent ideal laboratories within which to study both galaxy formation and dark matter. Understanding the effects of the environments they evolved in is important to producing accurate constraints on such physics. Present-day sensitivity limits have restricted the observational study of dwarf galaxies, especially the faintest ones, to the Local Group (LG; the Milky Way (MW), Andromeda, and all galaxies within $\sim$2 Mpc of each). Therefore, much effort has been put to studying MW-mass halos or LG-like pairs \citep[e.g.][etc.]{klypin1999, Springel2008, sawala2016, Applebaum2021} and the formation of dwarf galaxies within such environments \citep{Brooks2014, GK14elvis, wetzel2016LATTE, simpson2018, GK2017DM, GK2019LG, Buck2019, Akins2021}.

\lcdm~predicts the presence of dark matter (DM) structure and substructure at all scales, with galaxies populating halos down to a certain limit \citep{white1978,Moster2010,Jethwa2018}. Dwarf galaxies often exist within substructure of larger halos as well as being hosts to their own sub-substructure \citep{sales2013,wheeler2015}, with the Large Magellanic Cloud (LMC; the MW's largest satellite) as a well-studied example of this intermediate scale. Recent works investigating the observed and predicted satellite population of the LMC have revealed that these environments, while less disruptive than MW-like centrals \citep{Jahn2019}, can be host to significant populations of dwarf galaxies \citep{donghialake08, sales2011, wetzelDeasonGK2015, deason2015, jethwa2016magDES, sales2017, dooley2017,  kallivayalil2018MCsats, pardy2020, Erkal2020, SantosSantos2021}.

Environment -- the relative proximity to higher mass structures -- has been repeatedly shown to correlate with morphology, color, and star formation rate (SFR) in dwarf galaxies \citep{Balogh2004,Kauffmann2004,Baldry2006,Cooper2006,Lisker2007,Bamford2009,Wang2014}, though the majority of such surveys are limited to relatively massive environments and satellite galaxies with $M_\star \gtrsim 10^8$ \msun. Studies of LG and near-field dwarf galaxies have further corroborated this trend by demonstrating the dependence of HI abundance and star formation history (SFH) on environment \citep{Grebel2003,Grcevich2009,Weisz2011}. More detailed studies of the LG have shown that a majority of satellite dwarf galaxies in the LG are quenched at stellar masses below $10^8$ \msun~ \citep{Weisz2015, Wetzel2015quenching}, while field dwarfs remain star-forming \citep[][though this work is mostly limited to bright dwarf galaxies with $M_\star \gtrsim 10^7$\msun]{Geha2012}. In contrast to the universally high quenching fractions found around the LG, \citet{Mao2021} showed via the Satellites of Galactic Analogs (SAGA) survey that MW-mass systems outside the LG may have systematically lower quenched fractions, even at lower masses. This suggests that the SFHs of LG satellites might not be typical of MW-mass hosts. 

Plausible physical mechanisms for the quenching of star formation have been identified. Small-scale hydrodynamic effects are known to either remove gas from a star-forming galaxy via ram-pressure stripping \citep{Gunn1972,Abadi1999}, prevent the infall of cold gas that fuels star formation leading to `starvation' \citep{Larson1980}, and/or disrupt the structure of the galaxy during close interactions with other galaxies \citep{Moore1996,Pearson2016}. 

Since dwarf galaxies have shallow potentials and massive hosts such as the MW are know to host hot gaseous halos \citep{Gupta2012}, much attention has been given to the effects of ram-pressure stripping in low-mass satellites. For example, \citet{Fillingham2015,Wetzel2015quenching} demonstrated a mass-dependent quenching model in which satellites with $M_\star=10^{6-8}$ \msun~have short quenching time-scales consistent with ram-pressure stripping, while the longer quenching time-scales of higher mass satellites ($M_\star\gtrsim10^8$ \msun) are consistent with starvation. \citet{Fillingham2016} further showed that a clumpy gaseous halo with local densities $\sim 2 - 20$ times the mean gas density increases the efficacy of ram-pressure stripping and reproduces the high quenched fraction of MW satellites.  This model breaks down at the lowest galaxy masses ($M_\star \lesssim 10^5$ \msun), where ram-pressure and starvation have been shown to be unable to reproduce the universally early quenching times \citep{Emerick2016,RodriguezWimberly2019}, thus pointing to heating from the ionizing UV background as the quenching mechanism at such scales. 

While the quenching fraction of LG satellites is known to rise as satellite mass falls \citep{Weisz2011,McConnachie2012,Wetzel2015quenching}, high resolution studies of nearby satellites and simulated LG-like environments have enabled the characterization of the full star formation histories (SFHs) of dwarf galaxies, and the study of their dependence on both satellite mass and environment. \citet{Weisz2015} characterized a bimodal mass-dependence for quenched fraction, with the highest ($M_\star \sim 10^{11.5}$ \msun) and lowest ($M_\star < 10^5$ \msun) mass galaxies holding high quenched fractions, and galaxies with $M_\star = 10^{8-10}$ \msun~having the lowest, suggesting this mass range may be the most difficult to quench. Comparing to infall time estimates from \citet{Rocha2012}, they also find that higher mass satellites tend to quench $1 - 4$ Gyr after infall, while lower mass satellites quench prior to infall. \citet{wetzel2016LATTE} used FIRE-2 simulations \citep{hopkins2018fire2} to reproduce properties of satellites around MW-like hosts, in particular, the wide scatter in SFHs and the stellar mass dependence thereof. \citet{GK2019SFH} looked at a sample of $\sim$ 500 dwarf galaxies from the FIRE suite to investigate the effect of various environments (LG vs MW vs centrals thereof vs highly isolated centrals), finding that LG and MW-like environments quench similarly, and form their stars earlier than dwarf centrals, supporting the host-satellite interaction model for quenching. They also find that higher mass dwarf galaxies are more likely to form a higher fraction of their stars at later times, in agreement with observed LG SFHs. 

Other simulations of MW and LG-like environments, such as APOSTLE \citep{Digby2019}, Auriga \citep{simpson2018} and NIHAO \citep{Buck2019} have demonstrated consistent findings, particularly that satellites form earlier than centrals. Recently, \citet{Akins2021} used the DC Justice League simulation suite to show that ram-pressure is the source of short quenching time-scales for satellites of intermediate mass ($M_\star = 10^{6-8}$ \msun), as well as the diversity of satellite SFHs and the trend of increasing quenched fraction with decreased \mstr.

While interactions with the host environment are known to affect satellite star formation, it can also affect morphology through gravitational interactions. Such interactions can be strong enough, depending on the satellite's physical size, proximity of the host, and the mass of the host, to produce observable stellar features known as tidal streams. Tidal streams around our Galaxy have been studied extensively, starting with the Sagittarius (Sgr) dSph tidal stream \citep{Ibata1994,Belokurov2006}, and with dynamical models of stream kinematics revealing multiple close encounters of satellites with the MW \citep{Johnston1995,Majewski1996,Helmi2001,Penarrubia2005}. Such features have also been observationally identified in galaxies beyond the LG \citep[e.g. ][]{Malin1997,MartinezDelgado2008,MartinezDelgado2010}. Stellar streams give insight and evidence to the hierarchical nature of galaxy assembly. Given this hierarchical nature and the confirmation of a population of satellites around the LMC, tidal streams should presumably be detectable around galaxies of lower mass than the MW. A handful of tidal streams have indeed been discovered around dwarf galaxies \citep[][]{MartinezDelgado2012,Carlin2019}, but their cosmological frequency remains unknown.

Much of the literature on satellites and the interactions with their host environments is confined to the scale of the MW/LG. This is because our highest resolution observations of dwarf galaxies exist within this volume. As next-generation surveys such as DELVE \citep{Drlica2021} MADCASH \citep{carlin2016madcash} and LBT-SONG \citep{Davis2021} come online, it will be important to characterize the environments of lower-mass systems. We aim to extend the analysis of previous works listed above to the scale of the Large Magellanic Cloud (LMC), about an order of magnitude smaller than the MW. For example, \citet{Carlin2021} recently discovered two ultrafaint dwarf satellites of LMC-mass hosts approximately 3 Mpc from the MW. Previous works have characterized properties of LMC analogs \citep[][]{chan2015,chan2018,elbadry2018} or the predicted satellite population \citep{jethwa2016magDES,sales2017, kallivayalil2018MCsats, Jahn2019, pardy2020, Erkal2020,SantosSantos2021}, but limited work has been done on characterizing the influence such environments have on observable properties of the satellite population, in particular their SFHs, quenched fractions, and tidal structures. 

This paper is organized as follows: the simulations and sample are presented in Section \ref{sec:sims}; overall trends in satellite quenching are investigated in Section \ref{sec:sats_cens}, while we investigate the environmental quenching of individual satellites in Section \ref{sec:EnvQuenching}; tidal features around our LMC-mass hosts are presented in Section \ref{sec:tides}. 

\begin{table*}
\centering
 \begin{tabular*}{\textwidth}{l @{\extracolsep{\fill}} c c c c c c c c c} 
 Simulation & $m_\text{bary}$ & $M_{\text{200m}}$ & \mstr~ & $r_{\text{200m}}$ & r$_{50\star}$ & min. \mstr & N$_\text{satellite}$ & N$_\text{central}$ & Reference \\ [0.5ex]
	& (\msun) &(\msun) & (\msun) & (kpc) & (kpc) & (\msun) & & \\
 \hline

\texttt{m11c} & 2100 & 1.5e11  & 8.2e8 & 167.4  & 2.80 & 5.89e4 & 2 & 10 & 1  \\
\texttt{m11d} & 7070 & 2.8e11  & 4.1e9 & 203.9  & 6.75 & 2.29e5 & 6 & 1 & 2 \\
\texttt{m11e} & 7070 & 1.5e11  & 1.4e9 & 166.0  & 3.31 & 2.46e5 & 3 & 1 & 2 \\
\texttt{m11h} & 880  & 1.8e11  & 1.1e8 & 174.0  & 1.44 & 1.25e4 & 10 & 5 & 2  \\ 
\texttt{m11q} & 880  & 1.5e11  & 3.4e8 & 168.7  & 2.71 & 1.23e4 & 5 & 9 & 1 \\ 
\texttt{m11v} & 7070 & 2.9e11  & 2.4e9 & 210.5  & 2.61 & 1.39e5 & 4 & 5 & 1 \\
\hline
 & & & & & & \textit{total} & 30 & 31 \\

\end{tabular*}
\caption{Properties of the host halo of all FIRE simulations analyzed. Column 1 is the name of each run; column 2 is the minimum baryonic particle mass; column 3 (\mtwo) is the mass of DM contained within \rtwo; column 4 ($M_\star$) is the stellar mass of the primary central of each zoom-in region; column 5 (\rtwo) is the radius at which the mean interior DM density is equal to 200 times the average matter density of the universe; column 6 ($r_{50\star}$) is the half mass radius of stars in the primary central galaxy; column 7 (min $M_\star$) is minimum stellar mass of any object examined in each run (an effective resolution limit); column 8 is the number of satellites around the primary central; column 9 is the number of resolved galaxies outside \rtwo~of the primary central; and column 10 is the reference in which the simulation was first presented. The total count for satellites and centrals analyzed herein is shown in the bottom row. Satellites are luminous (sub)-halos located within the host \rtwo~at $z=0$, while centrals are located outside the host halo. We make further resolution cuts on contamination by low-resolution particles ($M_\text{lowres}/M_\text{200m}$ < 3 per cent) and a minimum of 20 star particles formed within the progenitor halo of each object. Naturally, the runs with higher particle mass are less capable of resolving low-mass galaxies, leading to incompleteness of the faint end. Note that \texttt{m11d}, \texttt{m11e}, and \texttt{m11v} are unable to resolve ultrafaints ($M_\star < 10^5$ \msun) when applying our resolution criteria.  References: [1] \citet{hopkins2018fire2}; [2] \citet{elbadry2018}  }
\label{tab:props}
\end{table*}

\section{Simulations}
\label{sec:sims}

We analyze six cosmological zoom-in simulations of isolated LMC-mass halos from the Feedback In Realistic Environments project\footnote{\url{http://fire.northwestern.edu}} (FIRE). These runs used the FIRE-2 model \citep{hopkins2018fire2} via the cosmological hydrodynamics code \texttt{GIZMO}\footnote{\url{http://www.tapir.caltech.edu/~phopkins/Site/GIZMO.html}} \citep{hopkins2015gizmo}, a multi-method gravity plus hydrodynamics code, in its meshless finite-mass (MFM) mode. 
GIZMO implements an improved version of the Tree-PM solver from GADGET-2 \citep{springel2005} with fully-adaptive conservative gravitational force softenings for gas \citep{price2007}.

Simulations are initialized\footnote{\url{http://www.tapir.caltech.edu/~phopkins/publicICs/}} with second-order Lagrangian perturbation theory at $z=99$ using the \texttt{MUSIC} code \citep{hahn2011music} and evolved within a low-resolution cosmological box. The intended ``zoom-in'' \citep{katzwhite1993,onorbe2015} volume is selected as a convex Lagrangian region containing all particles within ${\sim}5\,$\rtwo\ at $z=0$ with no similar or higher mass halos as the primary, and is then reinitialized with higher resolution. This procedure is iterated until convergence at the intended resolution, with a buffer of low-resolution particles surrounding the main volume. The simulation is then evolved until $z=0$. Since our hosts are isolated LMC-mass halos that are not embedded within the environment of the LG, small deviations in the assembly history and satellite population may be expected in comparison to the real LMC. However, the physical mechanisms explored here, in particular the environmental effects of LMC-mass hosts, are expected to apply in the case of the real LMC in addition to influence from its evolution in proximity to the LG environment.


The FIRE-2 code calculates heating and cooling rates from 10 -- 10$^{10}$~K, using \texttt{CLOUDY} ionization states for free-free, photoionization \& recombination, Compton scattering, photoelectric, metal-line, molecular, fine structure, dust collisional, uniform cosmic ray heating, against a spatially uniform, redshift-dependent UV background \citep{faucher2009}.

Star particles are formed in gas that is required to be locally self-gravitating, self-shielded, Jeans unstable, and with density $n_H > n_\text{crit} = 1000$ cm$^{-3}$; inheriting mass and metallicity from their progenitor gas particles. To calculate stellar feedback, a \citet{kroupa2001imf} IMF is assumed in each star particle, with feedback quantities tabulated from the STARBUST99 stellar population model  \citep{leitherer1999}, including supernova Type Ia, II, and stellar winds, as detailed in \citet{hopkins2018fire2} and \citet{hopkins2018sne}. Radiative feedback is modeled using the Locally Extincted Background Radiation in Optically-thin Networks (LEBRON), accounting for absorbed photon momentum, photo-ionization, and photo-electric heating.

Dark matter (DM) particles are assigned to halos and subhalos through 6+1 dimensional phase space analysis via the \rockstar~halo finder \citep{behroozi2013rockstar}, which determines gravitationally bound particles and assigns mass through a spherical overdensity calculation relative to a threshold, such as the critical density density of the universe. Stellar properties are computed for each subhalo during an iterative post-processing procedure in with star particles within 80 per cent of a halo's radius and slower than 2$\times$\vmax~ with respect to the halo center are selected and refined until the stellar mass converges to $<1$ per cent. More details on this process can be found in \citet{samuel2020}. Progenitors of $z=0$ halos are traced through time using \texttt{consistent-trees} \citep{Behroozi2013_trees} to construct merger trees. We use properties generated by the above methods to initially determine the stellar mass and star formation histories of dwarf galaxies, but we make further cuts on resolution as described below.

\subsection{Selecting the Sample}
Table \ref{tab:props} lists the simulations used in our sample, including the resolution, the halo and stellar masses of the host galaxies, as well as the number of satellites and centrals identified within each simulation volume.  Satellites are identified as being within \rtwo~of the main host halo at $z=0$, where \rtwo~is defined as the radius within which the mean DM density is equal to 200 times the average matter density of the Universe. We classify centrals as any galaxy that falls outside \rtwo~but within the high-resolution region of each simulation. Well-resolved centrals are further selected as having a maximum contamination of low resolution particles at 3 per cent of their $z=0$ halo mass. This cut is unnecessary for satellites, because they naturally fall within the high-resolution region. 

\begin{figure*}
    \centering
    \includegraphics[width=0.49\textwidth]{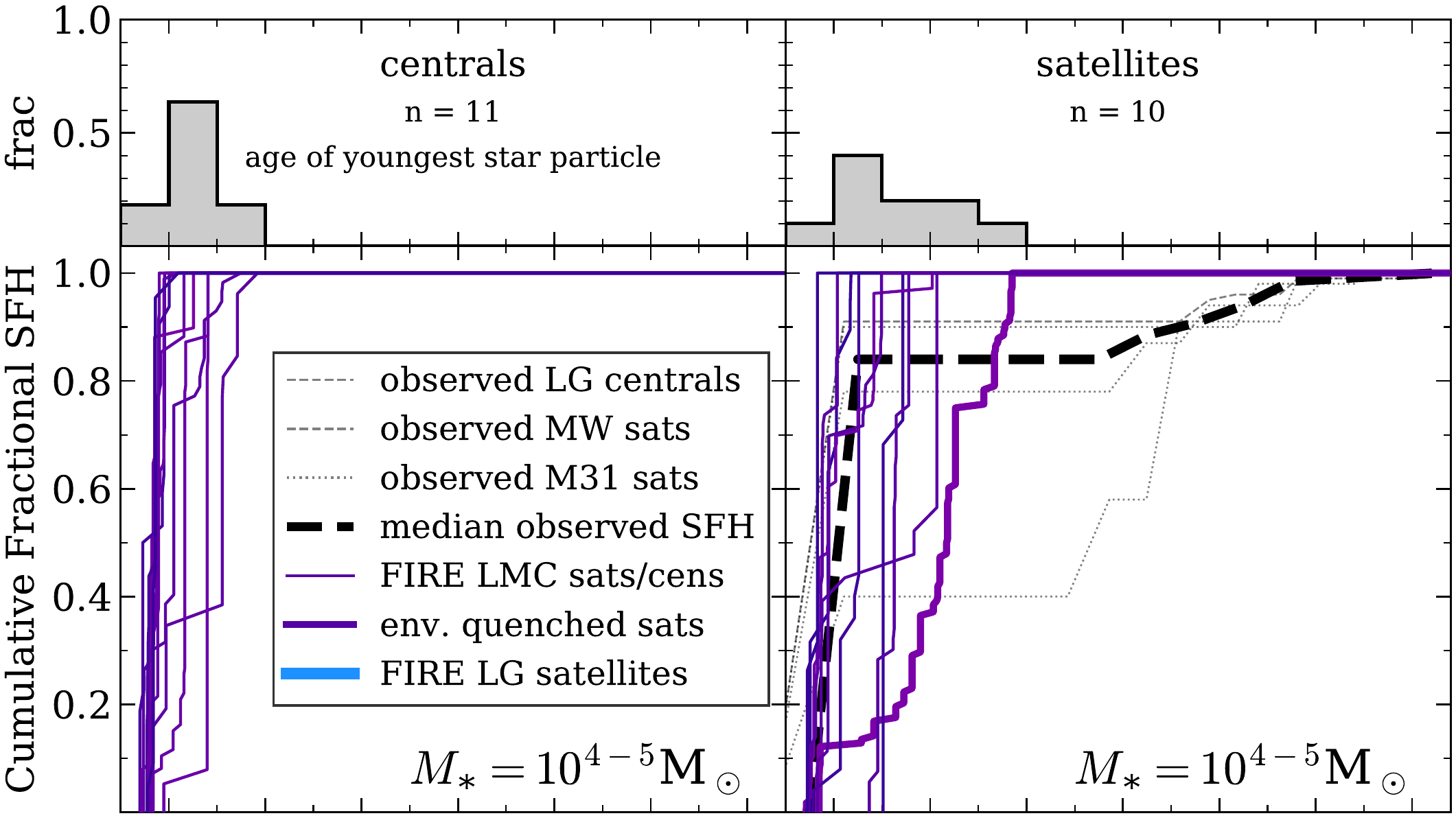}
    \includegraphics[width=0.49\textwidth]{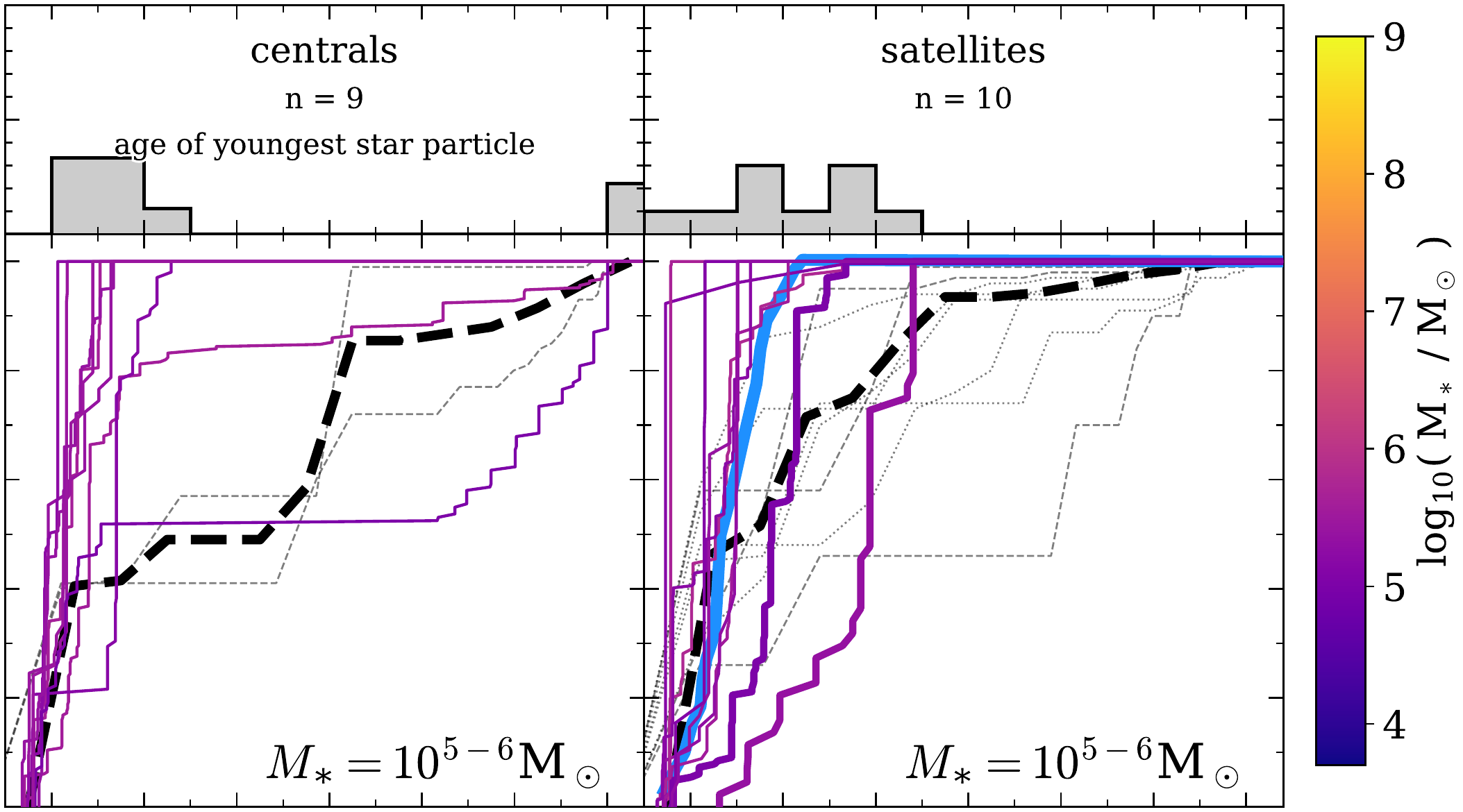}\\
    \includegraphics[width=0.49\textwidth]{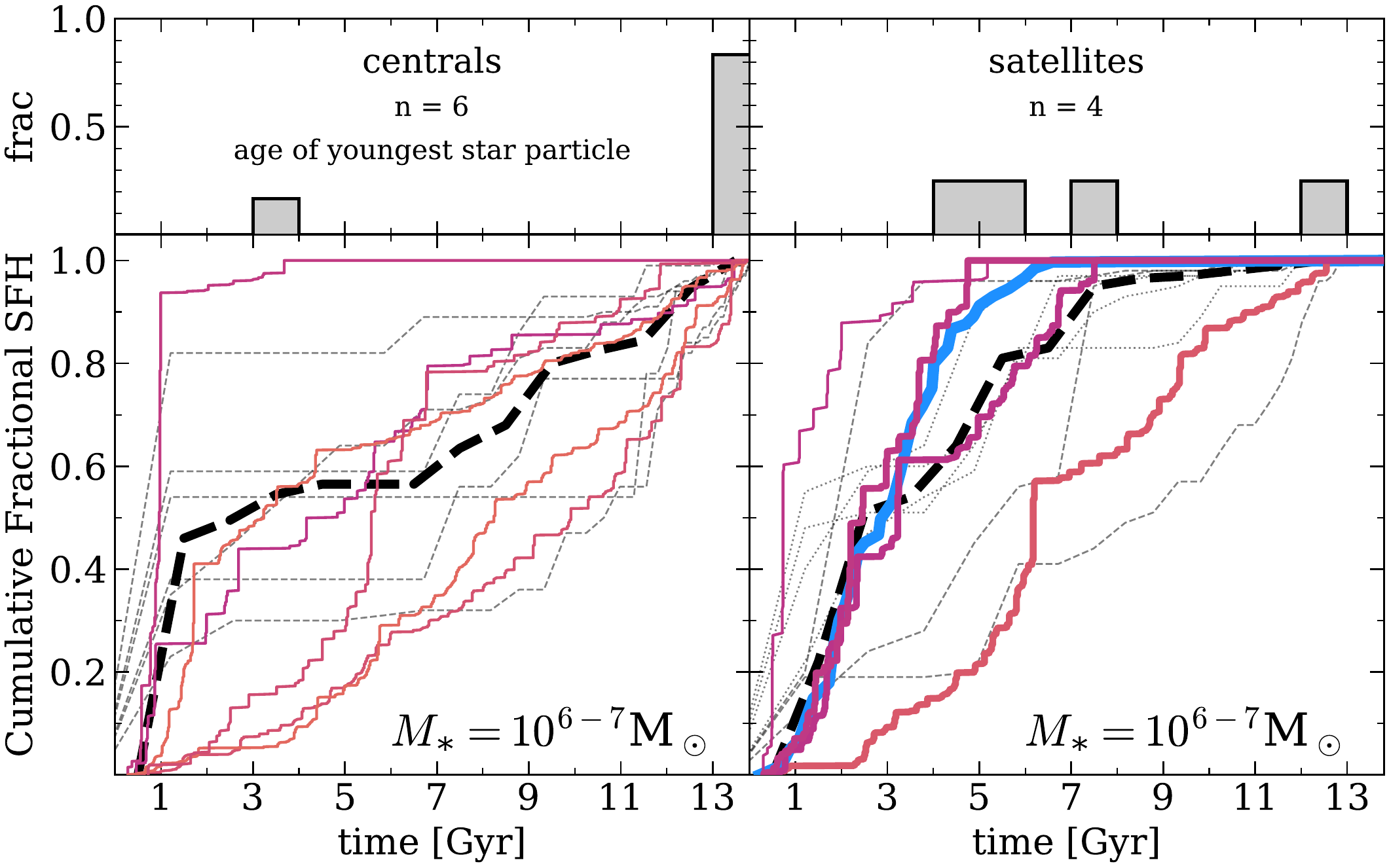}
    \includegraphics[width=0.49\textwidth]{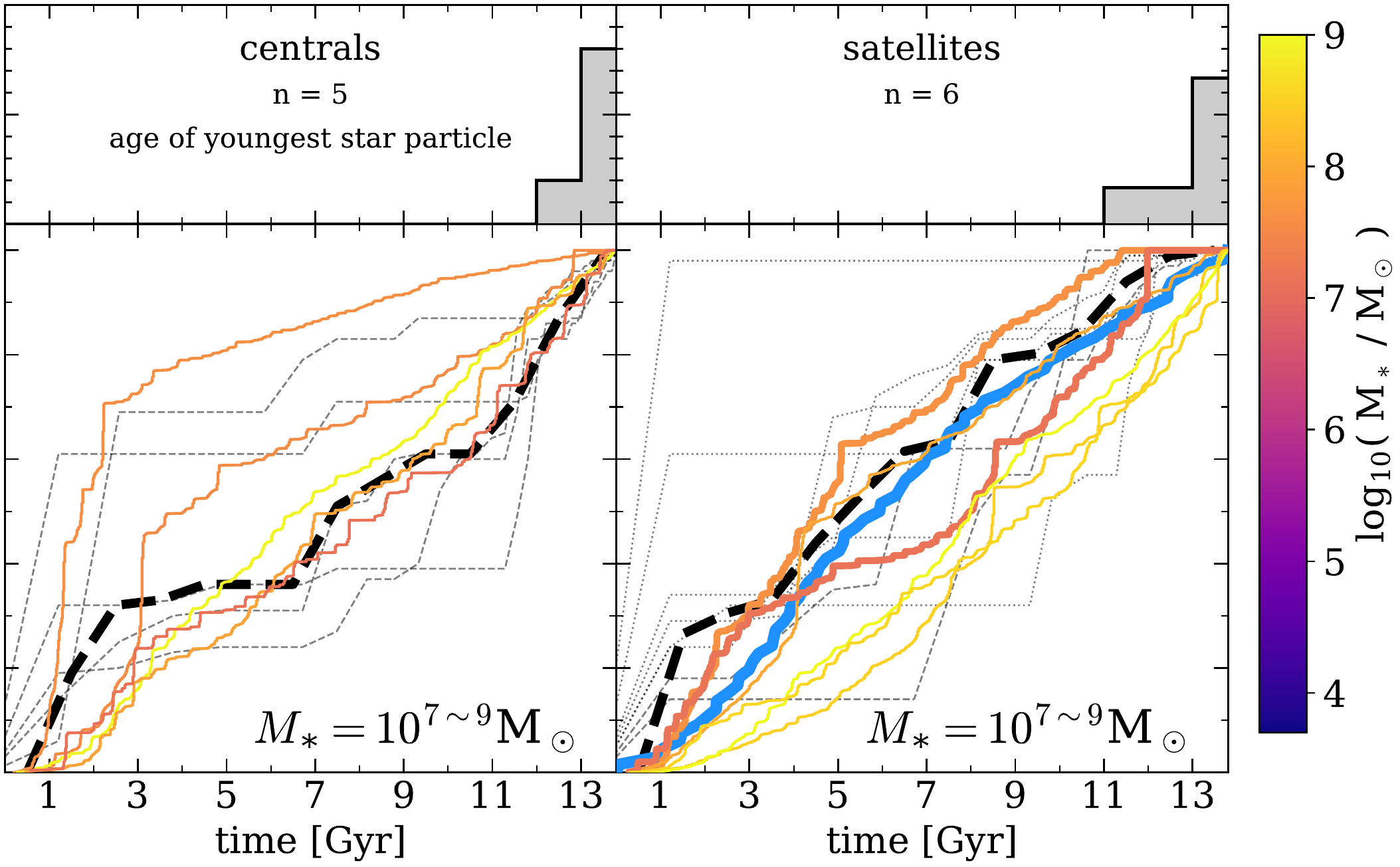}\\
    
    \caption{Star formation histories (SFHs) of simulated LMC satellites (right sub-panels) and centrals (left sub-panels), colored and separated according to stellar mass. Satellites identified as being environmentally quenched (see Section \ref{sec:EnvQuenching}) have been highlighted with thicker line styles. All colored lines are simulated data, while black/grey lines are observational data. Note that all observed data is for Local Group (LG) satellites/centrals. Individual observed SFHs of satellites of the MW/M31 \citep{Weisz2014,Skillman2017} are shown as faint dashed lines (style according to legend), with medians in each bin shown as thicker dashed black lines.
    Blue lines represent the median SFHs of satellites of LG-like hosts simulated using the FIRE-2 code, and as presented in \citet{GK2019SFH} (see Table 1 therein for more information, including resolution). Histograms on the top of each panel represent the formation time of the youngest star particle for simulated LMC satellites or nearby centrals (i.e. they do not include observed satellites of the MW/M31, nor simulated FIRE satellites of LG-like hosts).
    The majority of centrals are quenched prior to $t = 4$ Gyr if they are not presently star-forming. Satellites, however, are more prone to influence by the environment of their LMC-mass hosts, and exhibit a greater diversity of quenching times. Overall, the SFHs of LMC satellites do not differ substantially from the SFHs of LG satellites. This is perhaps a counter-intuitive result given the reduced stellar mass to halo mass ratio of LMC-mass halos, and their less disruptive nature \citep{Jahn2019}.}
    \label{fig:SFH}
\end{figure*}

To study the SFHs and to remove spuriously assigned particles, we track the location of star particles assigned to each (sub)halo and select only the ones which were formed within half of the halo's radius (of bound DM particles, as determined by \rockstar). We place a minimum cutoff of 20 such star particles for each galaxy analyzed. In some cases, merger events led to large amounts of stars formed outside the halo being assigned at later times, but as these are physically meaningful associations, they were retained. 

We have explicitly checked for splashback galaxies -- those that entered \rtwo~of the primary central at some point, but exited at a later time -- and find that all but one such objects are satellites at present day. Note that this is in contrast with the splashback population of MW-mass hosts, which tend to be significant outside of \rtwo~at $z=0$ \citep{Sales2007,GK14elvis,Fillingham2018}, an effect which is likely due to the difference in halo mass and infall rates of low-mass galaxies. The splashback process can potentially affect the evolution of the object and contaminate the sample of centrals, which are intended to represent the evolution of similarly-massed galaxies in FIRE that are not affected by the host (or at least affected to a substantially smaller degree). The individual splashback central is a low-mass galaxy (\mstr$\approx$3\e{5} \msun) that was quenched at $t\approx1.2$ Gyr while it was $\sim$1000 kpc from the central, first entering the primary halo at $t\approx11$ Gyr, suggesting that its quenching was not environmentally induced. Our entire population of centrals is therefore unlikely to have been directly influenced by the environment of the primary LMC-mass halo.

We note that the following predictions are limited by both resolution and sample size. Only three of the runs are able to resolve galaxies with $M_\star<10^5$ \msun, leading to a limited variety of cosmological volumes being sampled at this mass scale. Indeed, at all resolved scales, six zoom-in hosts is not a sufficient sample size to make statistically robust predictions, especially due to the relatively small number of predicted satellites of LMC-mass halos compared to that found around MW/LG-mass hosts. We consider this to be a case study in the formation of LMC satellites, rather than a statistically complete cosmological prediction. 

\begin{figure}
    \centering
    \includegraphics[width=0.45\textwidth]{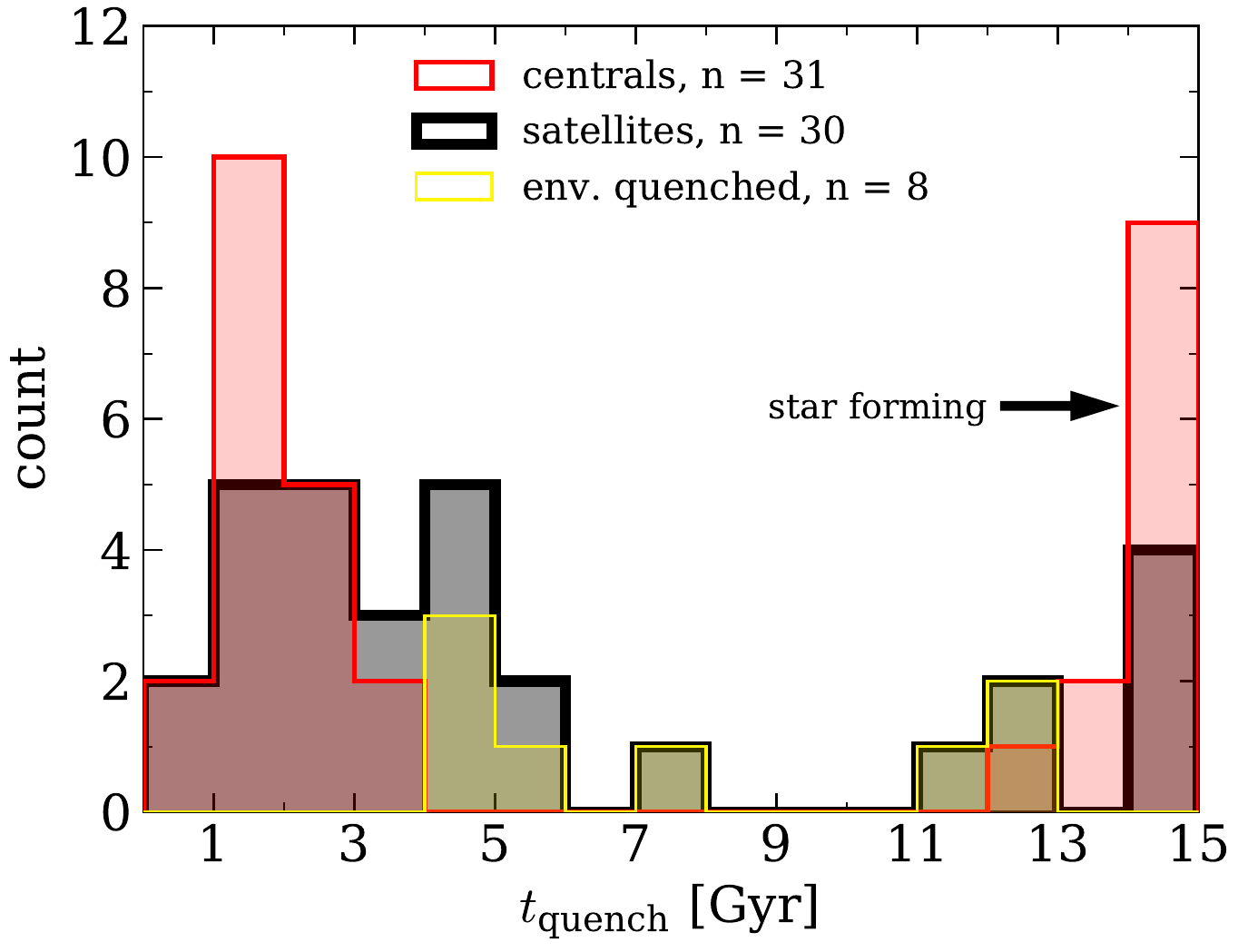}
    \caption{Histogram of quenching times for all centrals and $z=0$ satellites of LMC-mass hosts. Here, we define \tquench~as the formation time of the youngest star particle associated with the galaxy. Environmentally quenched satellites are highlighted in yellow and selected as having 4 Gyr $<$ \tquench $<$ 13.2 Gyr, as well as quenched within twice the virial radius of their host. To distinguish galaxies that were recently quenched from those that are actively star-forming, we have placed galaxies that formed a star particle within the last 500 Myr in the $14-15$ Gyr bin, labeled as `star forming.' While both centrals and satellites follow bimodal distributions, centrals exhibit clearly defined peaks and low scatter, while satellites exhibit a wider distribution. This indicates that the environments of LMC-mass hosts are the source of additional quenching mechanism(s), which we investigate in Section \ref{sec:EnvQuenching}.} 
    \label{fig:tqhist}
\end{figure}

\section{Satellite Quenching}
\subsection{Comparison to Centrals}
\label{sec:sats_cens}

Figure \ref{fig:SFH}~shows the SFHs of dwarf galaxies separated by association (satellites of an LMC mass host versus nearby centrals) as well as separated by stellar mass bins. All simulated data is shown as solid lines, while observed data for satellites of the MW/M31 is shown via dashed lines (these objects are shown for comparison, and are not direct analogs of our simulated satellites of LMC-mass hosts). We highlight LMC-satellites that have been identified as environmentally quenched (see Section \ref{sec:EnvQuenching}) in a thicker line. In this subsection, we examine only the objects from our sample of LMC-mass hosts, with individual SFHs shown as solid colored lines. Satellites are defined as being located within its host virial radius at $z=0$, while centrals are defined as being located outside this radius at $z=0$ but within the high-resolution region of the simulated zoom-in volume.  We combine the $M_\star=10^{7-8}$ \msun~and $M_\star=10^{8-9}$ \msun~bins due to the low number count, late quenching times, and general similarity of SFHs of galaxies in these bins. 

In every mass bin, satellites (right panels) exhibit a wider range in quenching times than do centrals (left panels). We find that all ultrafaint (UF; $M_\star \approx 10^{4-5}$ \msun) galaxies quench early on, as expected from the heating effects of the ionizing UV background, preventing the accretion and subsequent cooling of gas in low-mass halos. The FIRE-2 simulations implement a spatially uniform UV background, so such effects would not be due to patchy reionization \citep[][]{hopkins2018fire2}. The latest forming UF satellite (\tquench $\sim$ 4.6 Gyr) has been identified as environmentally quenched, and inhabiting a more massive halo than other satellites (when comparing the highest mass ever achieved by each halo to account for mass loss due to tidal stripping). This galaxy would have likely continued forming stars and be included in a higher mass bin if star formation were not shut off due to environmental effects. Excluding this object, we find that UF satellites quench at $t=2.0\pm0.8$ Gyr, and centrals quench at $t=1.3\pm0.6$ Gyr, using the age of the last star particle formed as quenching time. When looking at the 90 per cent star formation time-scale, we find that satellites (again excluding those that are environmentally quenched) have $\tau_{90} = 1.8\pm0.7$, and centrals have $\tau_{90} = 1.3\pm0.5$. We therefore find the distribution in quenching times between UF satellites and centrals to be statistically indistinguishable. We also note that UF dwarf galaxies are only present in the three highest resolution runs: \texttt{m11c}, \texttt{m11h}, and \texttt{m11q}.

In the next mass bin, we find that satellites with $M_\star=10^{5-6}$ \msun~also exhibit a wider range of quenching times than their predominantly early-quenched central counterparts. Notably, there are two late-quenching centrals, which formed stars until $t\approx 13$ Gyr. These galaxies inhabit somewhat larger DM halos (\mtwo$\approx$5.5\e{9} \msun) than most other galaxies in this stellar mass range, with the average halo mass excluded these two objects being \mtwo$\approx$2\e{9} \msun. This difference might seem small, but leads to a factor of 2 difference in their virial temperatures (4.6\e{4}K versus 2.3\e{4}K), increasing the temperature limit of gas that will remain bound to the halos during heating due to reionization and star formation feedback. The latest forming satellites in this mass bin, as in the previous, were identified as environmentally quenched and as possessing more massive DM halos than the rest of the satellites in this bin. Therefore, these objects are more consistent with failed versions of centrals in the next stellar mass bin.

Moving up in stellar mass to $M_\star = 10^{6-7}$ \msun, we now find that the majority of centrals are star-forming (the exception again being an outlier in halo mass, this time much lower than average), while all satellites are quenched at various intermediate times, \tquench$\approx4.5-12.5$ Gyr. {\it This is a strong indication that the environment of LMC-mass hosts is able to quench star formation in its dwarf companions}. 

In contrast, the highest mass dwarf satellites ($M_\star = 10^{7 \sim 9}$ \msun) form universally late, with all centrals and four of six satellites remaining star-forming at $z-0$. The two quenched satellites ceased forming stars at $t\approx11-12$ Gyr, and have stellar masses of $M_\star = 10^{7-8}$ \msun. All galaxies with $M_\star > 10^8$\msun~are star-forming, regardless of environment, further supporting the claim that such galaxies are the most resilient to quenching \citep[][]{Wheeler2014,Weisz2015,Wetzel2015quenching,Fillingham2015}.

Figure \ref{fig:tqhist} shows the histogram of quenching times for all satellites and centrals. We define star-forming galaxies as those which have formed at least one star particle in the last $\sim$ 500 Myr, while quenched galaxies did not form any star particles in the same time interval. Quenching times for such objects are defined as the formation time of their youngest associated star particle. We include the count of star-forming galaxies in the $t=14-15$ Gyr bin to differentiate this population from those that formed their last star particle between $t=13-13.2$ Gyr. We find that centrals demonstrate strongly bi-modal quenching behavior, either halting their star formation by $t=4$ Gyr, or continuing until $z=0$. In contrast, satellites exhibit a wider variety of quenching times, with 8 having clear signs of environmental quenching (highlighted in yellow). These galaxies were selected with 4 Gyr $<$ \tquench $<$ 13.2 Gyr, and $d_\text{host}(t=t_\text{quench}) < 2$\rtwo$(t)$. That is, they are selected as having quenched late enough that reionization heating is unlikely to be the culprit, and close enough to their host halo to be influenced by its circum-galactic medium (CGM). We investigate these objects further in Section \ref{sec:EnvQuenching}. 

In summary, we predict that isolated LMC-mass galaxies should host a population of mostly quenched low-mass dwarf galaxies. More specifically, these galaxies should be host to $\gtrsim 3$ ultrafaint ($M_\star = 10^{4-5}$ \msun) satellites with ancient stellar populations, $1 - 4$ intermediate mass ($M_\star = 10^{5-7.5}$ \msun) dwarf satellites with a variety of quenching times, some of which may have been environmentally quenched by the host, and lastly, one bright star-forming companion of $M_\star \gtrsim 10^{7.5}$ \msun, though not all simulations contain such an object. 
The number and mass distribution is in agreement with recent predictions of the LMC satellite population \citep{jethwa2016magDES,sales2017,kallivayalil2018MCsats,pardy2020,Jahn2019,Erkal2020}. 

\begin{figure}
    \centering
    \includegraphics[width=0.4\textwidth]{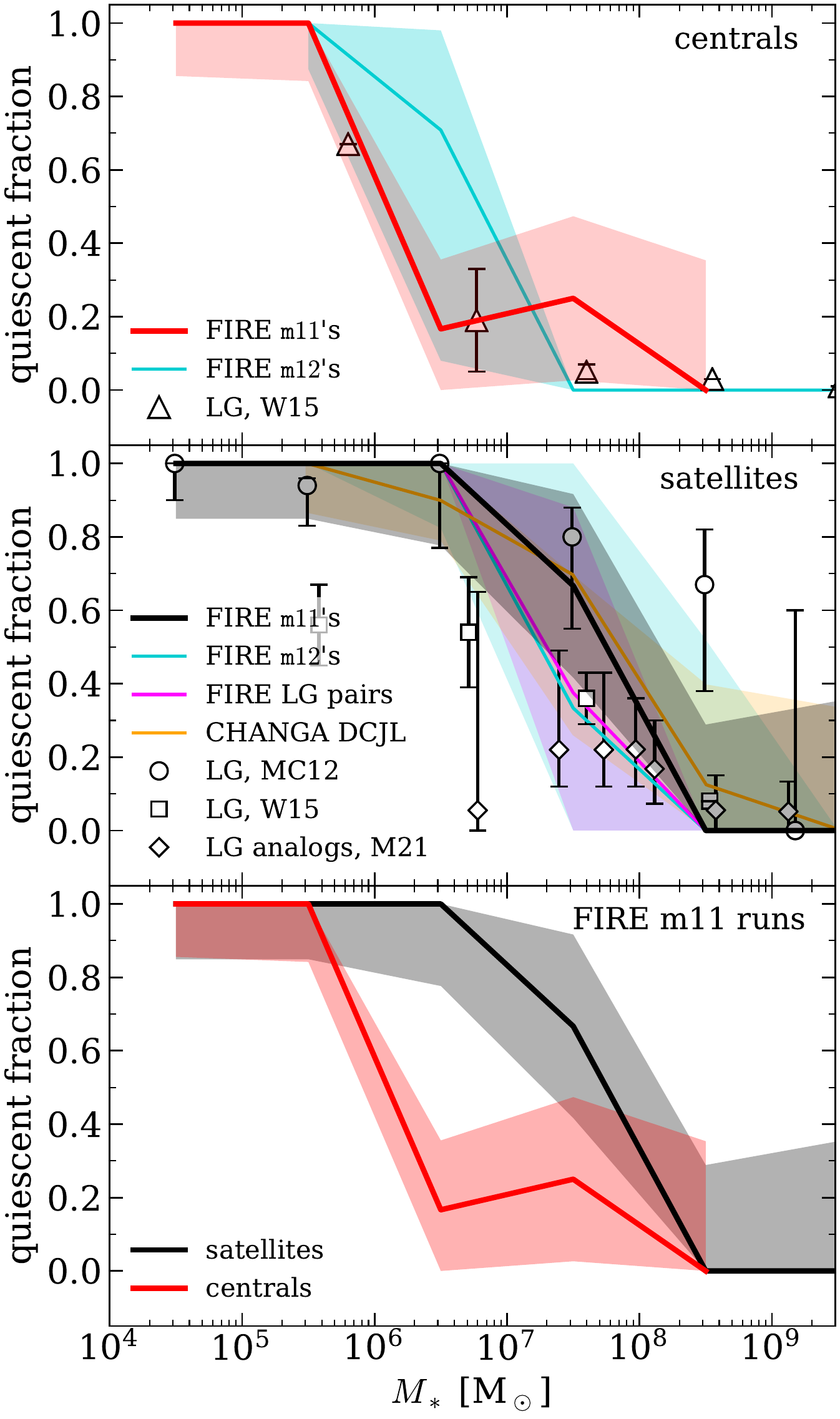}
    \caption{Stellar mass versus quiescent fraction in our simulations of LMC-mass hosts versus observed (points) LG galaxies and simulated (shaded regions) LG or MW-like halos. MC12 \citep{McConnachie2012} and W15 \citep{Weisz2015} report data for the LG itself, while M21 \citep{Mao2021} report data for observed MW analogs. We define the quiescent fraction as the number of galaxies in each mass bin that have \tquench $<13.2$ Gyr (that is, selecting all galaxies that have not formed a star particle within the last 500 Myr) to allow for variations that may arise from finite time-steps and star particle mass limits. We find that LMC-mass hosts are able to quench their satellite population to nearly the same degree as MW or LG-like hosts, perhaps a surprising result given their significantly smaller halo masses and baryonic content. References: FIRE MW+LG - Samuel et al. \textit{in prep}; CHANGA - \citet{Akins2021}; MC12 - \citet{McConnachie2012} (as compiled by \citealt{Wetzel2015quenching}); W15 - \citet{Weisz2015};  M21 - \citet{Mao2021}}
    \label{fig:mstar_fquench}
\end{figure}

\subsection{Comparison to Local Group Environments}
We find that satellites of LMC-mass hosts quench their satellites similarly to the LG, with simulated data from the FIRE simulations for such objects from Figure 4 of \citet{GK2019SFH} shown as the blue line in the right panel for each mass bin of Figure \ref{fig:SFH} (where $M_\star = 10^{4-5}$ \msun~is not included due to resolution limits). These simulations of LG-like environments include two MW-mass  ($\sim10^{12}$ \msun) halos, and satellites are defined as being with 300 kpc of one of those MW-like halos. For our $M_\star = 10^{7-9}$ \msun~bin, we show the mean reported SFH for the bins $M_\star = 10^{7-8}$ \msun~and $M_\star = 10^{8-9}$ \msun. In each mass bin where simulated LG-satellites are available, the overall shape of SFHs for such objects are consistent with LMC satellites. 

We also compare to observed SFHs for satellites of the MW or M31 \citep{Weisz2014,Skillman2017}, with individual SFHs in each mass bin shown as thin dashed/dotted lines, while the median is shown as a thicker black dashed line.
We find that our simulated LMC satellites have SFHs that are broadly consistent with observed MW/M31 satellites at fixed stellar mass, especially in the two highest mass bins. The observed galaxies in the lowest bins tend to form later than their simulated counterparts. This is perhaps due to observational uncertainty, since the majority of stars in each observed galaxy are formed in early times, consistent with our simulated galaxies. Constraining the exact time of quenching can be a challenge with observational data. For example, see \citet{Weisz2015} for a discussion of the impact of blue straggler stars on the estimation of SFHs via color-magnitude diagram fitting. 

The similarity of SFHs between observed \& simulated LG/MW/M31 satellites and our simulated LMC satellites suggests that quenching of satellites may not be restricted to high-mass systems, and that dwarf-dwarf quenching could proceed likewise to quenching in LG-type environments, an effect that may be impactful on the interpretation of future observational missions categorizing satellites of LMC-mass hosts. There is evidence that the CGM of the MW is dense and structured enough to affect the evolution of its intermediate-mass dwarf satellites \citep[e.g.][]{Grcevich2009,Peek2009,Nakashima2018}. The apparent similarity in SFHs between Local Group and LMC satellites suggests this may be true of LMC-like environments as well.

Figure \ref{fig:mstar_fquench} shows the quiescent fraction (i.e., the portion of galaxies which have not formed a star particle within the last 500 Myr) versus stellar mass of LMC satellites and centrals along with additional observed and simulated LG satellites. Due to the fact that not all of our simulated LMC-mass systems contain satellites in each mass bin, our error bars are derived from Poisson scatter. Error bars on W15 were calculated from the difference in reported quenching fractions when considering morphological dTrans galaxies as either star-forming or quiescent. 

Largely, our population of LMC satellites quenches similarly to LG satellites. Consistent with observations of satellites within the LG \citep[][]{McConnachie2012,Wetzel2015quenching} as well as with simulated MW/LG satellites \citep[in FIRE-2 -- Samuel et al. \textit{in prep}; and CHANGA -- ][]{Akins2021}, we find that LMC satellites with $M_\star < 10^7$ \msun~are universally quenched, and satellites with $M_\star \gtrsim 10^8$ \msun~are predominantly star-forming. This is also in agreement with semi-analytical models of the LG population \citep{Fillingham2016, Fillingham2019}. The interim region of $10^7$ \msun$ < M_\star < 10^8$  \msun~consists of satellites that are either presently star-forming or quenched within the last $\sim$2 Gyr. In contrast, we find that 90 per cent of nearby centrals are quenched below $M_\star < 10^6$ \msun, while 91 per cent of centrals with $M_\star > 10^6$ \msun~are star-forming by $z=0$, with outliers in quenching status also being outliers in halo mass. This indicates that satellites of $M_\star = 10^{6-7}$ \msun~are ideal probes of  environmental quenching, while satellites with $M_\star > 10^8$ \msun~are difficult to quench, in agreement with previous work on the quenching of satellites of higher mass hosts \citep{Wheeler2014,Weisz2015,Fillingham2015, Wetzel2015quenching}. We therefore find that LMC-mass hosts, though they possess a reduced amount of substructure, may in fact be able to quench their satellites in a similar manner as MW/LG-like environments.

\begin{figure}
    \centering
    \includegraphics[width=0.46\textwidth]{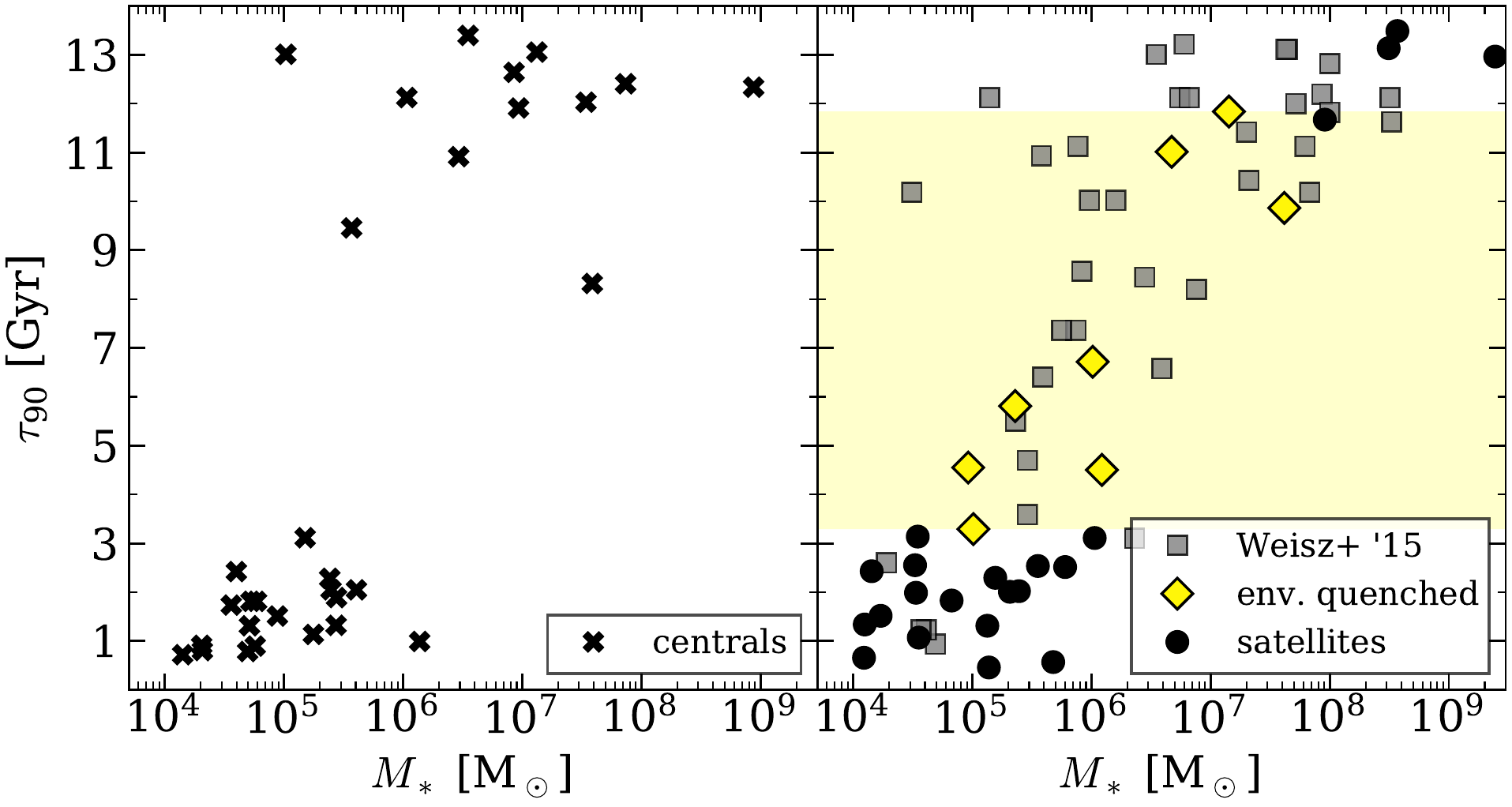}\\
    \includegraphics[width=0.47\textwidth]{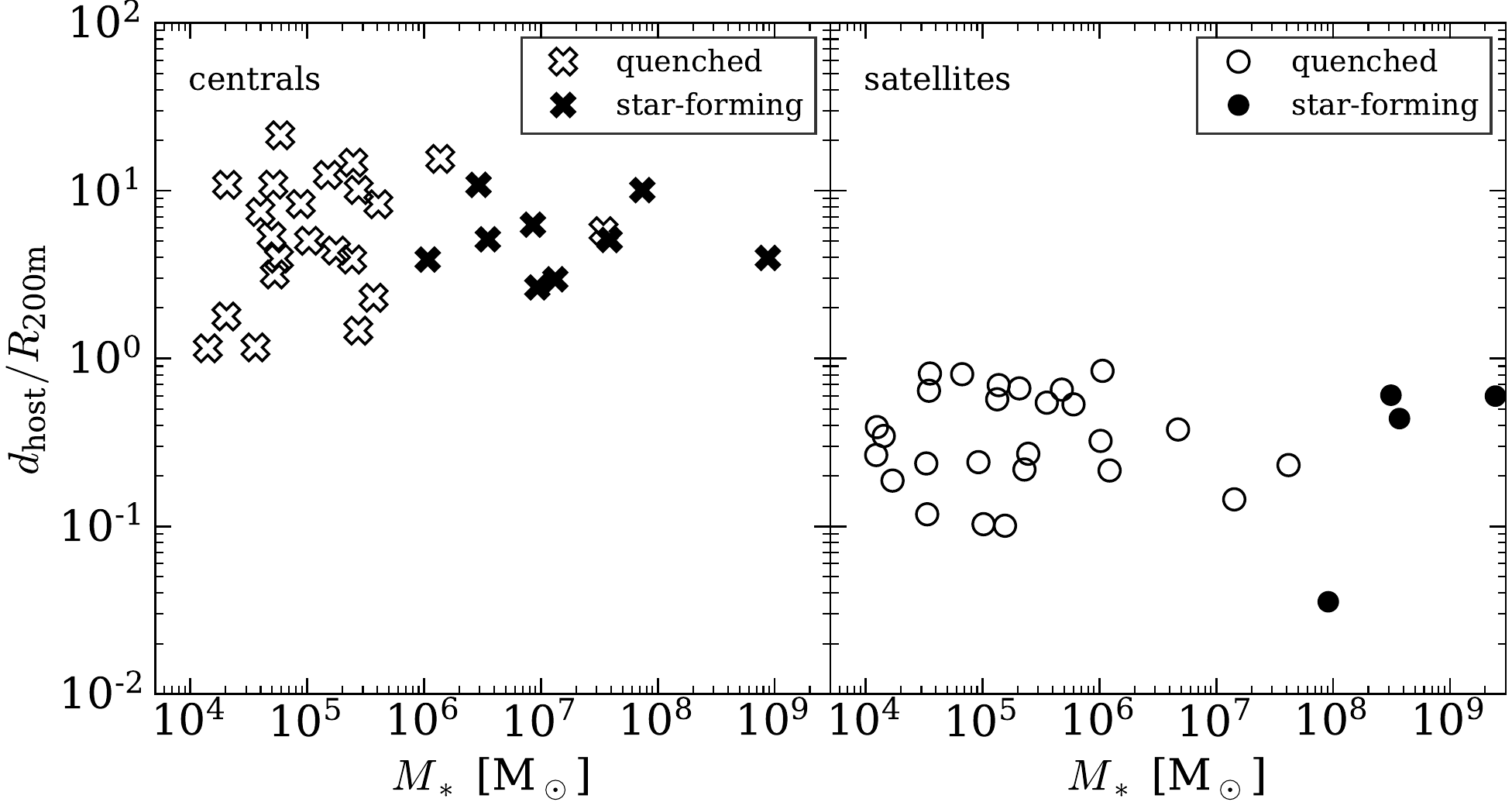}
    \caption{(\textit{top panels}) Stellar mass versus the 90 per cent star-formation time-scale ($\tau_{90}$) for simulated satellites and field galaxies, including observed local-group dwarf galaxies from \citet{Weisz2015}. 
    We use $\tau_{90}$ here as a comparable analog to the observational data, representing the cosmic time at which 90 per cent of the present-day stellar mass was formed. The yellow shaded region denotes the range in $\tau_{90}$ for environmentally quenched satellites, which are plotted as yellow diamonds (see text for identification criteria). 
    We find that our simulated environmentally quenched satellites fall within the distribution of observed galaxies in the local group. We find fewer low-mass late-forming satellites, though that could be due to the high variance of quenching times and our small sample size. (\textit{bottom panels}) Stellar mass versus $z=0$ distance to the host galaxy, normalized by \rtwo. Star forming galaxies are shown as solid markers, while quenched galaxies are shown as open markers. }
    \label{fig:mstar_tau90}
\end{figure}

Figure \ref{fig:mstar_tau90} shows the stellar mass and 90 per cent quenching time-scale ($\tau_{90}$) for our simulated centrals and satellites, as well as observed LG satellites as reported by \citet{Weisz2015}. The highlighted region indicates the $\tau_{90}$ range of environmentally quenched satellites (see Section \ref{sec:EnvQuenching}). Note that $\tau_{90}$ and \tquench~are not identical quantities; \tquench~indicates the formation time of the youngest star particle, while $\tau_{90}$ represents the time at which 90 per cent of the $z=0$ stellar mass was formed. The environmentally quenched satellites of LMC-mass hosts are consistent with the trend of observed LG satellites in their stellar masses and $\tau_{90}$, further supporting the case that isolated LMC-mass hosts can environmentally influence their satellites similarly to the LG. The bottom panel of Figure \ref{fig:mstar_tau90} shows the $z=0$ distance to the primary central normalized to its \rtwo~versus stellar mass for both satellites and centrals, also marking their star-forming state. Note that there are far fewer star-forming satellites than centrals. Consistent with Figure \ref{fig:mstar_fquench}, we define quenched satellites as those with \tquench $<$ 13.2 Gyr (i.e. not having formed a star particle in the last $\sim$500 Myr).


\begin{figure}
    \centering
    \includegraphics[width=0.48\textwidth]{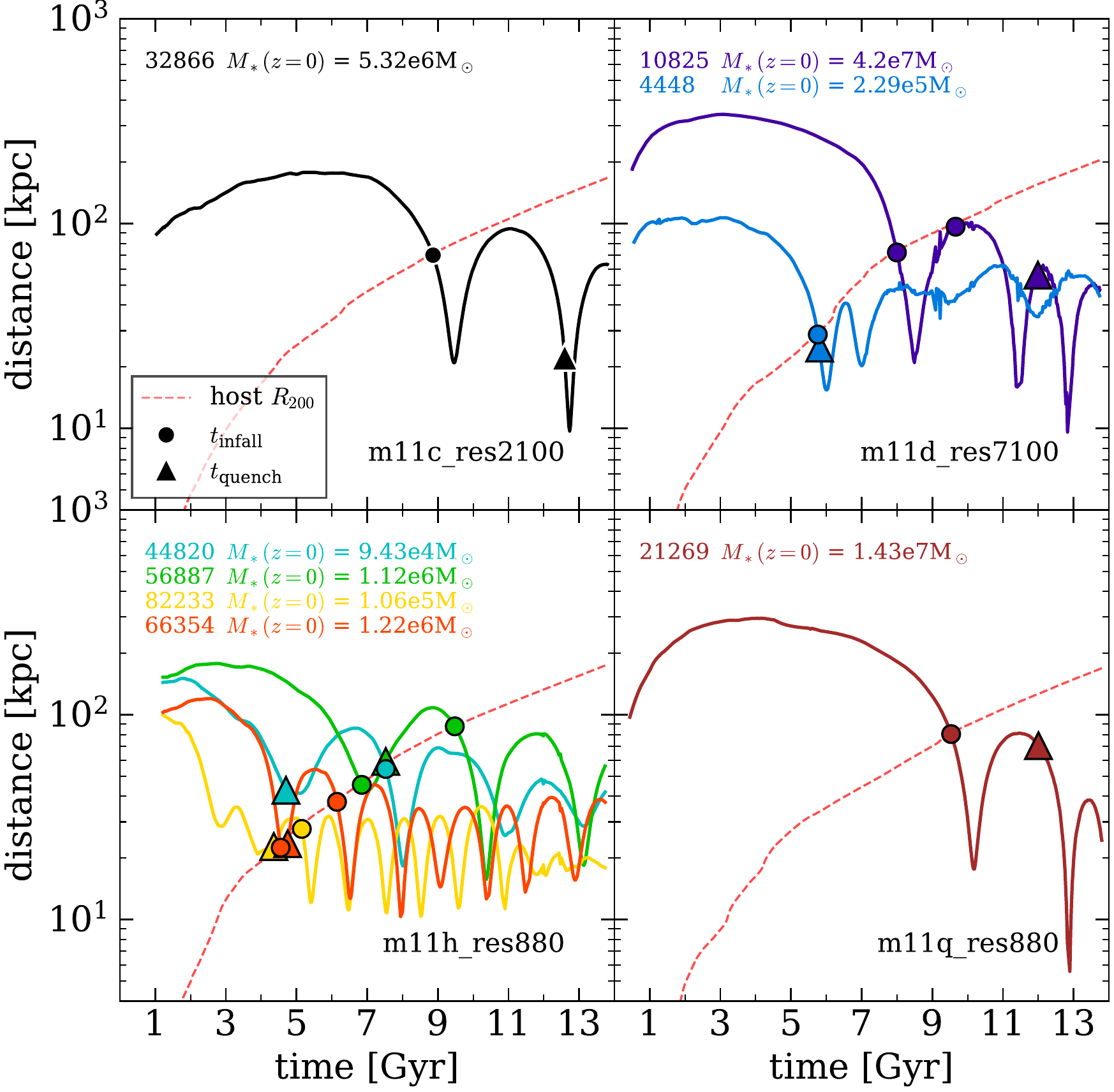}\\
    \hspace{3mm}
    \includegraphics[width=0.4\textwidth]{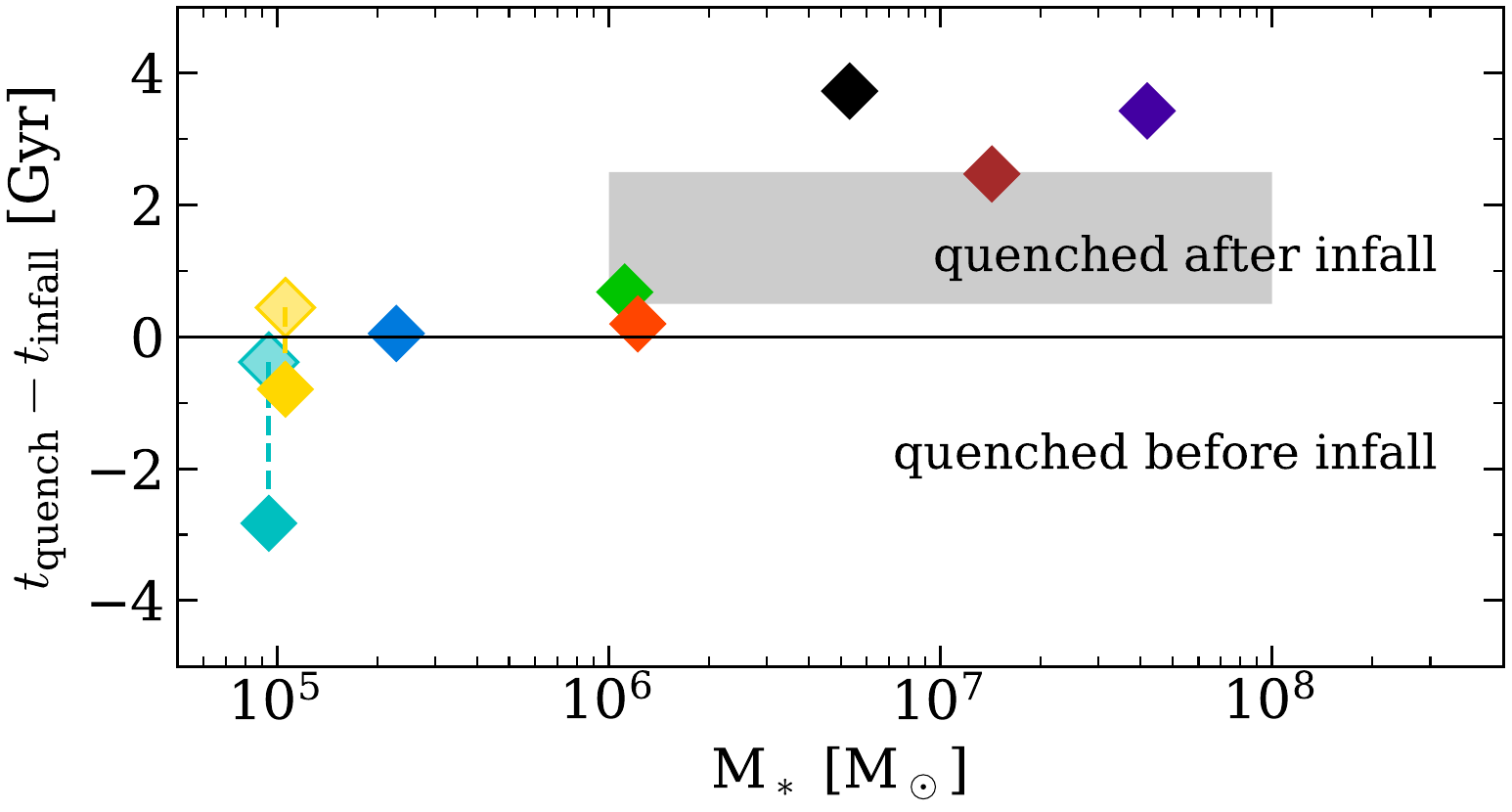}
    \caption{(\textit{top panels}) Orbits of environmentally quenched satellites (EQSs), with the virial radius of the host shown as a red dashed line, infall times marked as circles, and quenching times (i.e. the formation time of the youngest star particle) marked as triangles. This population of satellites was selected as having been quenched between 4 and 13 Gyr, and that were quenched at a distance of less than twice the virial radius of their host at that time. Five of the eight galaxies were quenched before th,eir first pericenter, suggesting the gaseous halos of these simulated LMC-mass hosts are rich enough to affect satellite evolution as far away as their virial radii. (\textit{bottom panel}) Quenching time-scales for EQSs, defined as $t_\text{quench} - t_\text{infall}$ such that galaxies which were quenched after infall appear above the horizontal line. The grey bar indicates the quenching time-scale due to stripping and feedback predicted for satellites of MW-like hosts \citep{Fillingham2016}. In addition, we plot the time-scales corresponding to the time of closest approach for subhalos 44820 \& 82233 (partially transparent cyan \& yellow markers) of \texttt{m11h} due to the fact that they come within the vicinity of the host halo around their respective quenching time, but splash back on wider orbits before later falling into the host halo.} 
    \label{fig:orbits}
\end{figure}

\subsection{A Closer Look at Environmental Quenching}
\label{sec:EnvQuenching}
Here we investigate the specific circumstances of quenching for the 8 identified environmentally quenched satellites (EQSs). Satellites were identified as being environmentally quenched by requiring intermediate to late quenching times such that 4 Gyr $<$ \tquench~ $<$ 13.2 Gyr, and proximity to the host halo  $d_\text{host}(t_\text{quench}) < 2r_\text{200m}(t_\text{quench})$. We allow for objects to be located outside the host virial radius at quenching time due to previous works highlighting the consistency of galaxies within 2\rtwo~of the MW with environmental quenching \citep{Fillingham2018}, indicating that the sphere of influence of the primary central is not strictly limited to such a radius. Three satellites quench outside their host \rtwo: 44820 at 1.7\rtwo, 56887 at 1.06\rtwo, and 82233 at 1.05\rtwo.

Figure \ref{fig:orbits} shows the orbits of these objects, as well as the evolution of the host virial radius, whose intersection with each orbit defines the infall times (marked as circles). Halo ID numbers are shown on the figure in corresponding colors, and consistent coloring will be used in further plots that highlight this sample. Five of the eight EQSs were quenched at or near the host virial radius, often with infall times shortly before or after their quenching times, suggesting that the circumgalactic medium (CGM) of the hosts are dense enough to influence satellites of this mass. These galaxies are lower mass, with stellar masses of $M_\star\approx 10^{5-6}$ \msun~and peak halo masses of $M_\text{halo,peak}\approx2\times10^9$ \msun. The other three EQSs are more massive, with $M_\star\approx10^7$ \msun, and $M_\text{halo,peak}\approx10^{10}$ \msun. These satellites fell into their host halos later, and quenched after first pericenter. 

This trend can be seen in the bottom panel of Figure \ref{fig:orbits}, which shows the stellar mass of EQSs compared to their quenching time-scales, defined as $\tau_\text{quench}  = t_\text{quench} - t_\text{infall}$. The three most massive EQSs have $\tau_\text{quench}\approx2-4$ Gyr, while the lowest mass EQSs have $-1$ Gyr $\lesssim \tau_\text{quench} \lesssim 1$ Gyr. Objects 44820 (cyan) \& 82233 (yellow) undergo a pericentric passage around the host before falling within \rtwo~(note the difference between quenching time and infall time markers on Figure \ref{fig:orbits}, and the orbital minima that occur near quenching time). We therefore include a secondary $\tau_\text{quench}$ based on the time of this pericenter rather than the infall time, as the boundary of the DM halo is somewhat arbitrarily defined, especially when considering the baryonic effects of the central galaxy. These points are shown as partially transparent markers connecting to the original point based on $t_\text{infall}$ via a dashed line, and bring them into stronger agreement with the other low-mass EQSs, with faster quenching time-scales such that $t_\text{infall} \approx t_\text{quench}$.

The distinction in quenching time-scales and infall times of low-mass versus intermediate-mass satellites suggests that there may be further stellar-mass dependence within the quenching model of \citet{Fillingham2016}. We have indicated their predicted quenching time-scales due to feedback and ram-pressure/turbulent stripping for satellites of MW-mass hosts in the bottom panel of Figure \ref{fig:orbits} as a grey bar. We find that intermediate-mass ($M_\star \approx 10^{6.5-7.5}$ \msun) satellites of LMC-mass hosts have somewhat longer quenching time-scales than predicted for satellites of MW-mass hosts. This makes sense in light of the lower stellar-mass to halo-mass ratio for dwarf galaxies like the LMC, and the predicted lower level of disruption for such systems when compared to MW-mass hosts \citep[][]{Jahn2019}. We also find that low-mass satellites ($M_\star \lesssim 10^6$ \msun), which are not resolved in the analysis of \citet{Fillingham2016}, have somewhat lower quenching time-scales than predicted for intermediate mass satellites. This is likely due to the lower binding energy of their less massive DM halos, leading to higher susceptibility to ram-pressure stripping and therefore quenching earlier in the infall process from less dense gas in the outer parts of the parent halo. In principle, this mechanism should apply to hosts of any mass, suggesting fast quenching time-scales (perhaps within -0.5 to 0.5 Gyr) for low-mass satellites of MW-like hosts. It is unclear at this point whether these two subtypes (i.e. $M_\star \approx 10^{6.5-7.5}$ \msun~with $t_\text{infall} > 8 $ Gyr and quenching time-scales of $2 - 4$ Gyr versus $M_\star \lesssim 10^6$ \msun~with $t_\text{infall} < 7 $ Gyr and quenching time-scales of -0.5 to 0.5 Gyr) lie on a continuous distribution of satellite quenching behavior, or if there is a stellar mass cutoff between distinct populations. 

\begin{figure}
    \centering
    \includegraphics[width=0.48\textwidth]{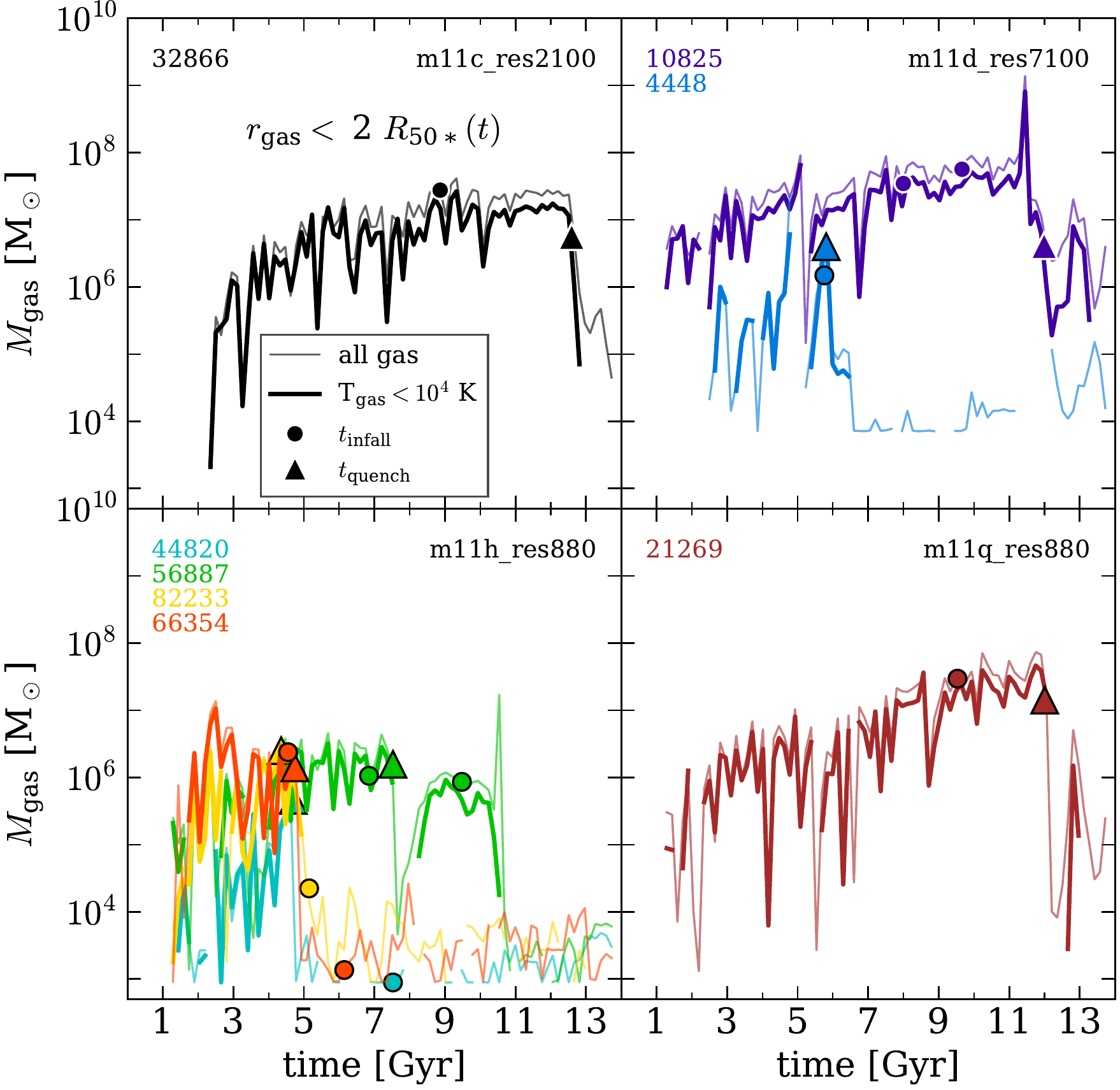}\\
    \includegraphics[width=0.48\textwidth]{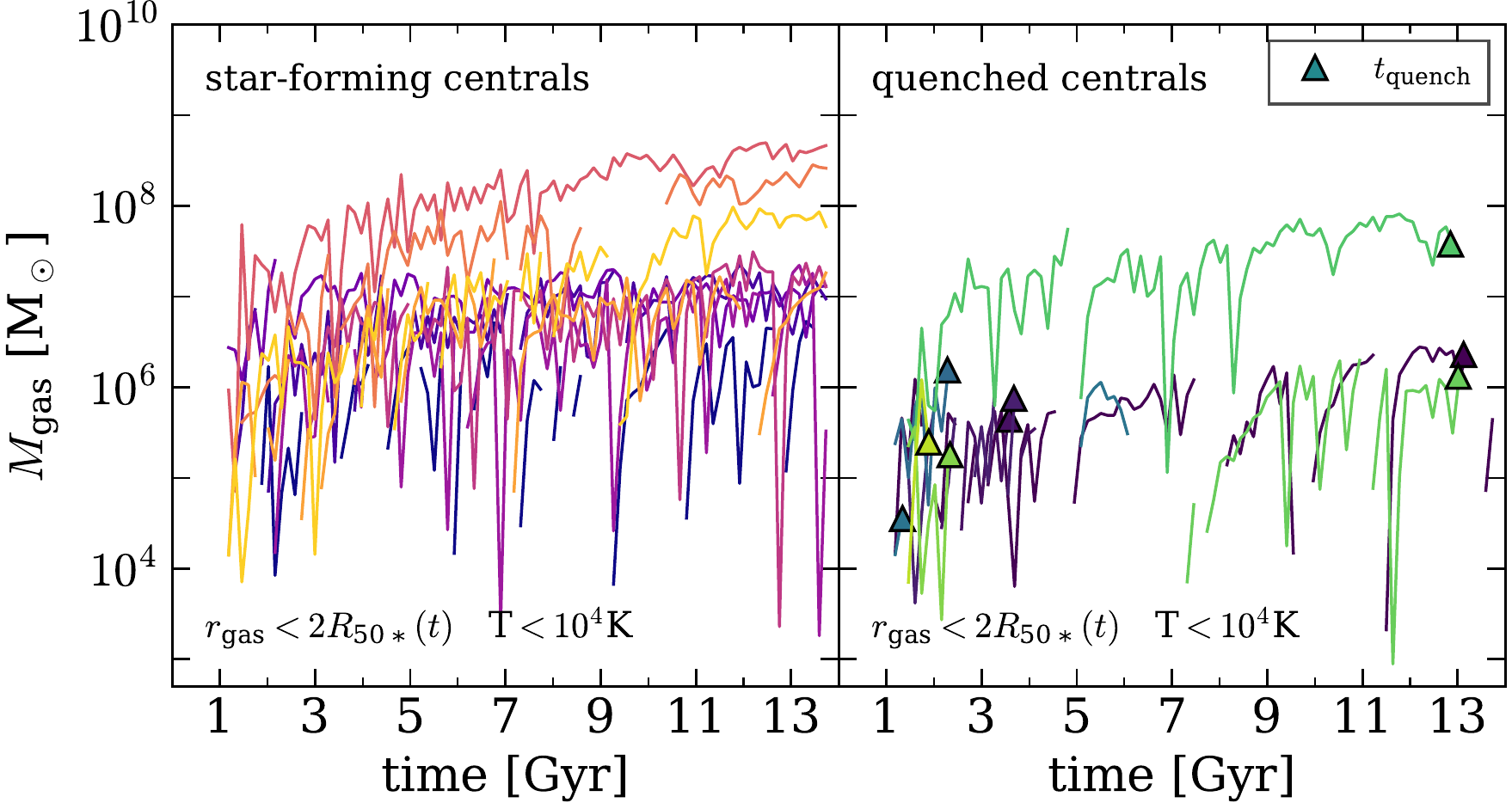}
    \caption{(\textit{Top 4 panels}) Total (thin) and cold (thick) gas mass within 2$R_{50\star}(t)$ for each EQS, excluding gas with high relative velocity to the satellite such that $|v_\text{gas}-v_\text{sat}| < 10\times$\texttt{max}$[v_\text{circ},\sigma_{v}]$, all measured within the satellite. Each galaxy experiences a steep drop in $M_\text{gas}$ at or near its quenching time, indicating a removal of gas through either star formation bursts or ram pressure (or both) rather than starvation or gravitational stripping, which are characterized by slower reductions in $M_\text{gas}$ (on the order of several Gyr or longer). (\textit{Bottom 2 panels}) Cold gas mass within 2$R_{50\star}(t)$ for star-forming (left) and quenched (right) centrals. We find a consistent presence of cold gas throughout the history of centrals that are star-forming at $z=0$, while quenched centrals cease to contain cold gas after their quenching times. } 
    \label{fig:mgas}
\end{figure}

\begin{figure}
    \centering
    \includegraphics[width=0.48\textwidth]{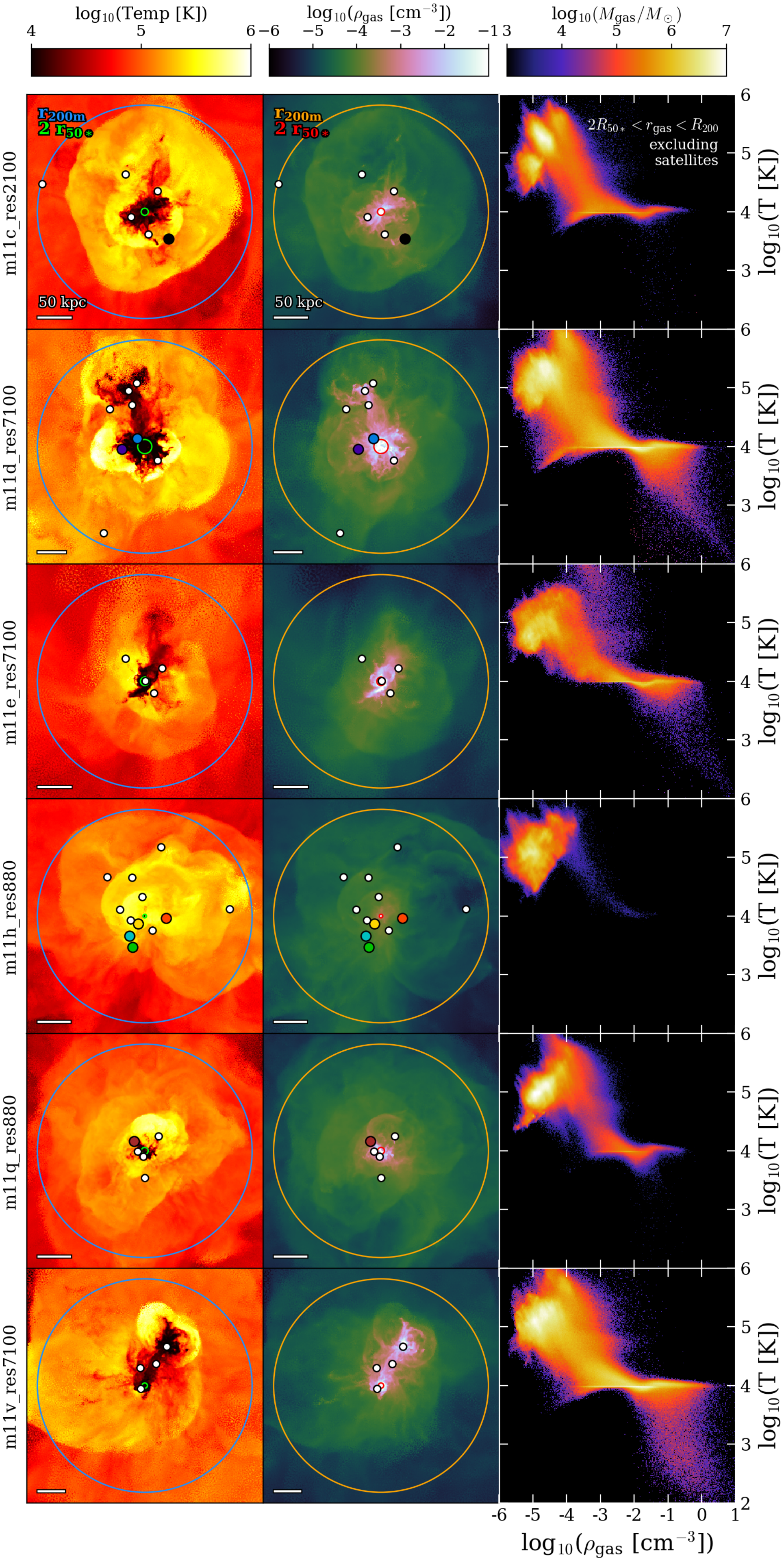}
    \caption{Properties of the gaseous halos of all LMC-mass halos at $z=0$. The left column shows a projection of gas temperature, the second column shows a projection of gas density (with points marking the locations of satellites), and the right column shows the phase-space diagram for halo gas, defined as $2R_{50\star} < r_\text{gas} < R_\text{200m}$ and outside $2R_{50\star}$ of luminous satellites. Each host exhibits a significant mass in a hot (T$=10^{4.5-6}$K) corona with $M = 3-6$\e{9} \msun, with highly structured regions of hot (rarefied) versus cold (dense) gas, with visible shock fronts.}
    \label{fig:cgm_panel}
\end{figure}

The top panels of Figure \ref{fig:mgas} show the total and cold gas mass within $2 r_{50\star}(t)$ for each EQS as a function of time. We find universally steep drop-offs in gas content near the quenching time for each galaxy, suggesting some form of hydrodynamic gas removal, which operates on much faster time-scales than gravitational stripping or starvation \citep[][]{Fillingham2015,Emerick2016}. 

While some galaxies retain or even re-accrete some amount of gas, none re-ignite their star formation after the initial gas-loss event. Take, for example, the \texttt{m11h} satellite 56887 (bottom left panel, green line), which is on a splashback trajectory before settling permanently in an orbit within the virial radius of the host halo at $t \sim 9$ Gyr. This satellite loses its gas and quenches after first infall, but is able to regain some gas on its trajectory back out of the host halo. It is possible that some or all of this gas is not tightly bound to the satellite, as our velocity cuts are somewhat liberal, but $\sim10^6$ \msun~of gas remains within that radius during the object's splashback orbit for another few Gyr before it infalls again and fully loses all remaining gas content. Interestingly, the re-accretion of gas to pre-quenching levels is not sufficient to reignite star formation in the satellite. The correlation of multiple infall and subsequent gas removal events is an encouraging suggestion that the environment of the host halo is responsible for stripping away any gas bound to the satellite. We therefore turn our attention to the CGM properties of the LMC-mass host galaxies.

The bottom 2 panels of Figure \ref{fig:mgas} show the history of cold gas within star-forming and quenched centrals. There are no obvious signatures in this data that distinguish the gas content of centrals from EQSs. We therefore look into further details of the gas content of EQSs in Section \ref{sec:ram_pressure}. For now, we turn our attention to the gas content of the host halos.

\subsubsection{Characterizing the Gaseous Halos of LMC analogs}
Figure \ref{fig:cgm_panel} shows the temperature and density projection for each LMC-mass host in out sample, as well as the phase diagram, with each pixel colored according to the total mass of gas contained within it. Projections are constructed by selecting gas with $|x| < 1.1$\rtwo, $|y| < 1.1$\rtwo, and $|z| < 0.2$\rtwo, where $z$ is perpendicular to the plane of the figure. We choose no particular orientation with respect to the host galaxy. This gas is then divided into evenly spaced 2-D ($x,y$) pixels, giving a column of gas with length 0.4\rtwo. The pixel is then colored according to the median physical value of temperature or density for all particles within its boundary. If there are no particles within the pixel, it is colored according to the median value of nearest non-empty pixels.

We find that a hot, richly structured gaseous halo is present around all LMC-mass hosts to varying degrees. While not all centrals are host to EQSs, the ubiquity of the rich gaseous halo suggests that the presence of such galaxies relies more on varying cosmological abundances of structure than it does on the ability any particular LMC-analog to quench its satellites. We identify two primary components of the CGM based on features in the phase diagrams: the hot corona, found in the upper left quadrant, and the horizontal feature of T $\sim 10^4$ K gas with $10^{-4} \lesssim \rho /$cm$^{-3} \lesssim 10^0$. Some runs also contain a small component of cold, dense gas in the lower right quadrant. Star-forming gas in FIRE is restricted to densities above $10^3$ cm$^{-3}$, and is not abundant enough outside of $2R_{50\ast}$ compared to the halo gas to appear on this figure. 

Quantifying the hot corona as gas with $10^{4.5} <$ T/K $< 10^6$, and 10$^{-6} < \rho /$cm$^{-3}< 10^{-4}$, as well as being located outside 2$r_{50\star}$ of the host galaxy and all satellite galaxies, we find that LMC-mass halos have 3$\sim$6\e{9} \msun~of gas in their hot coronas. Additionally, we find mean gas densities of $\sim$5\e{-4} cm$^{-3}$ and mean temperatures of $\sim$1\e{5} K, both quantities volume-weighted. These predictions are in good agreement with the detection of a hot ionized component in the LMC \citep{Wakker1998,Lehner2007} suggesting the presence of hot gas around the Magellanic clouds as well as with recent theoretical arguments of a need of a hot corona in the LMC to fully explain the morphology of the Magellanic stream \citet{Lucchini2020}. While there are differences in the presence and radial distribution of hot and cold CGM components between LMC and MW-mass galaxies in the FIRE-2 simulations \citep{Stern2021}, the existence of a relatively massive hot component out to $\gtrsim$ 100 kpc is consistent between host mass scales. 

\begin{figure*}
    \centering
    \includegraphics[width=0.85\textwidth]{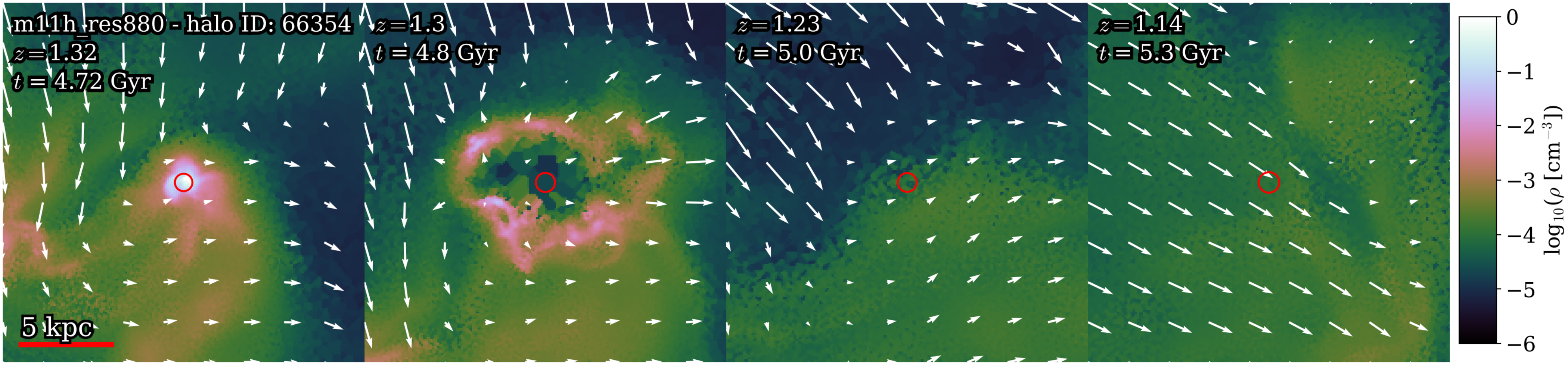}\\
    \includegraphics[width=0.85\textwidth]{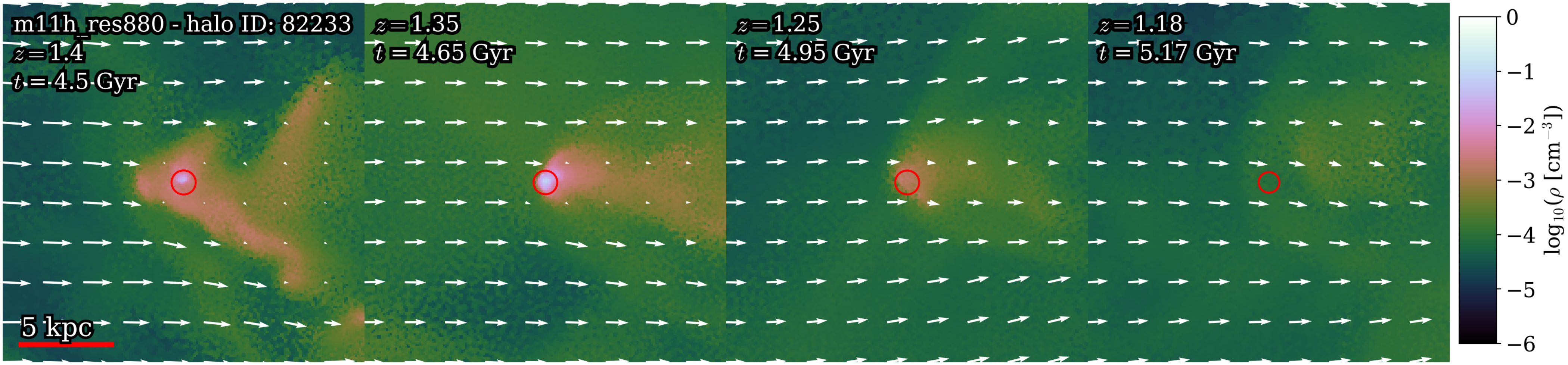}
    \caption{Gas density (color) and velocity (arrows; in the reference frame of the satellite) projection of two satellites of \texttt{m11h} as they fall into the host halo, demonstrating two primary modes of quenching: SF-driven stripping (\textit{top}) and standard ram pressure stripping (\textit{bottom}). Each image is centered on the halo location as provided by \rockstar~at the shown snapshot, despite the lack of obvious structure in later panels. The satellites are similar in size, with \mtwo(\tquench)2$\approx$3\e{9} \msun~and \mstr(\tquench)1$\approx$2\e{6} \msun, and they both quench around $t \sim 4.5$ Gyr. The first panel of both rows shows the satellite near \tquench, with visible tails due to its motion through ambient halo gas. The top row demonstrates quenching due to a combination of feedback and ram pressure, which is a common mechanism among our population satellites. The bottom row demonstrates a classic example of ram pressure stripping of a satellite.}
    \label{fig:gas_stripping}
\end{figure*}

\subsubsection{Quenching via Ram-Pressure and Feedback}
\label{sec:ram_pressure}


There are many possible sources for gas removal in satellites, for example, energetic feedback from star formation which can be induced by the increased pressure of the host environment, interactions with other galaxies, or ram-pressure stripping from the ambient halo gas. The time-scale of gas removal seen in Figure~\ref{fig:mgas} is short enough to rule out starvation, which occurs on longer time-scales as gas reservoirs within the satellite are depleted \citep{Fillingham2015}. Interactions such as fly-by events and ram-pressure stripping are functions of environmental properties (abundance of satellites, density of gas), while feedback-driven, self-induced quenching only depends on the star formation history of each galaxy (though the SFH may also be dependent on host environment). It is likely that a combination of these effects simultaneously occurs in orbiting satellites. 

Although the energetic feedback of the FIRE simulations is certainly enough to strongly affect the ISM of dwarf galaxies \citep[][]{elbadry2016,elbadry2017,elbadry2018,hopkins2018fire2}, the general lack of isolated dwarfs with \tquench $=4-13$ Gyr and $M_\star = 10^{5-7}$ \msun~makes self-quenching alone an unlikely cause for the halting of star formation in these satellites. However, one could not rule out environmentally induced starbursts (i.e. from compression of gas at orbital pericenters), or removal of low density gas blown out by feedback, which may have cooled and fallen back into the satellite if it were in isolation, but is easily swept away by the high density of the host's ISM. Such effects, which may not neatly be described as strictly environmental or strictly self-induced, seem to drive the evolution in some of the satellites in our sample, as illustrated by the two case studies presented here. 

Figure \ref{fig:gas_stripping} shows a series of density projections at four sequential time stamps of two low-mass satellites of \texttt{m11h} that were quenched near its virial radius. Also shown is the normalized gas velocity field in the reference frame of each satellite. Time stamps were chosen simply to highlight the state of the gas in and around each satellite as it is quenched, with the first panel being chosen as the snapshot immediately prior to the formation of its last star particle. The stellar half mass radius of each satellite is also shown as a red circle. Each frame is centered on the satellite's position at the given time.

The top row shows a satellite with \mtwo(\tquench) $\approx$ 3 \e{9} \msun, $M_\star(t_\text{quench}) \approx$ 2\e{6} \msun, and $M_\text{gas}(t_\text{quench}) \approx 10^6$ \msun, where \tquench~$=  4.75$ Gyr. It demonstrates trails characteristic of ram pressure in the first panel, but the gas is sufficiently dense in its core as to resist stripping. The velocity field reveals turbulence around the galaxy as well, though there is a clear front of gas moving downwards from the top of the figure. The second panel shows a burst of star formation that moves this gas out of the central region, heating and rarefying it. This enables the gas to be pushed out of the halo by the pressure from ambient halo gas in the third panel, resulting in no clear gaseous component to the halo in the fourth panel, where the velocity field has become more uniform. This process is generally consistent with ram pressure stripping, though it requires sufficient stellar feedback to `loosen' the gas within the satellite before the ambient halo pressure is capable of stripping and quenching it.

The bottom row shows a second similar mass satellite, with \tquench~$\approx$ 4.5 Gyr, this time with a much more uniform velocity field. This object demonstrates a more standard picture of ram pressure characterized by a gas stream extending from the satellite opposite the direction of motion. There is no feedback event that processes the gas prior to stripping - the pressure from ambient halo gas is sufficient to strip away the dense, bound gas within the satellite. Note that the second panel shows an increased amount of dense gas within $r_{50\star}$ due to compression via the ambient velocity field. The time-scale for each galaxy to go from possessing dense, concentrated, star-forming gas to possessing virtually no gas is $\sim 300$ Myr in both cases, though it is slightly faster in the case where feedback is involved.

An important qualification to this analysis is that both satellites come from the same parent halo -- \texttt{m11h}. This halo is host to an unusual abundance of satellites: 10 in total (12 including all subhalos with assigned star particles, forgoing the cuts described in Section \ref{sec:sats_cens}). As seen on the left side of the first panel of the top row in which an additional locus of dense gas is present, satellite-satellite interactions can also be a source of environmental quenching. This particular event seems to have compressed the gas in the satellite shown, leading to a strong burst of star formation, rarefying the gas and making it more susceptible to ram-pressure stripping via the halo gas. These objects were chosen for the case study due to their high resolution and obvious visual features. We have done a similar analysis of all EQSs and find ram-pressure alone or in combination with feedback from star formation to be the quenching mechanism for all EQSs.

Interactions can also be seen in the orbits of the above objects in Figure \ref{fig:orbits} which appear to have pericenter with some object other than the host prior to final infall. We have checked this explicitly, though the other satellites are not shown on the figure for visual clarity. We include this type of interaction under the umbrella of environmental quenching, though it does require the presence of sufficiently many companion galaxies for satellite-satellite interactions to take place. It is unclear how cosmologically common this is for LMC-mass hosts, but in our set of 6 centrals and 30 satellites, we identified 1 host with 2 instances of interactions. 

The pre-processing of satellites prior to infall is a natural prediction of \lcdm~\citep[][]{Li2008,wetzelDeasonGK2015,Benavides2020}, with part of the aim of this study to understand how the environment of the LMC could have affected its satellites prior to the group's infall into the halo of the MW. We expect that pre-processing -- whether due to prior group association or individual fly-by events -- before to infall into LMC-mass halos will perhaps be less common than for systems like the MW, simply due to the relative abundance of structure in each. However, this example demonstrates that pre-processing on much smaller scales than the MW is indeed possible, and perhaps contributes to the relatively high amount of environmentally quenched satellites within \texttt{m11h}.

\section{Effects of Tides on Satellites of LMC Analogs}
\label{sec:tides}
It has been shown that MW-mass galaxies are hosts to rich tidal features including coherent stellar streams and kinematically mixed stellar halos \citep{Helmi1999}. These features result from interactions between dwarf satellites on close orbits with their more massive hosts that tidally strip mass (both dark and luminous) from their companions. Similar processes are expected to occur also for satellites of lower mass hosts, with a handful of observations confirming the presence of tidal streams in satellites of dwarf-mass centrals \citep[e.g., ][]{MartinezDelgado2012}.

\begin{figure}
    \centering
    \includegraphics[width=0.35\textwidth]{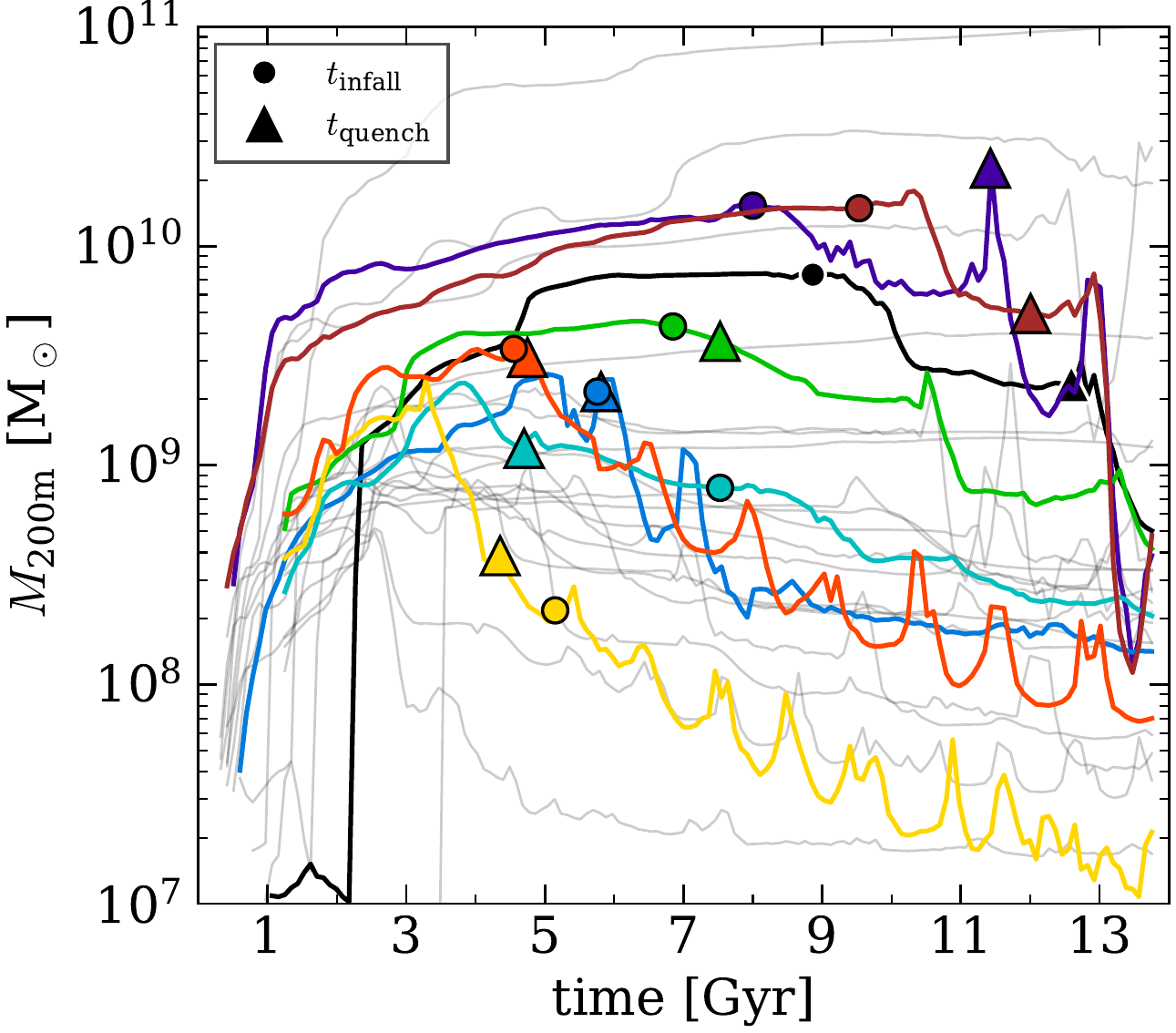}\\
    \includegraphics[width=0.35\textwidth]{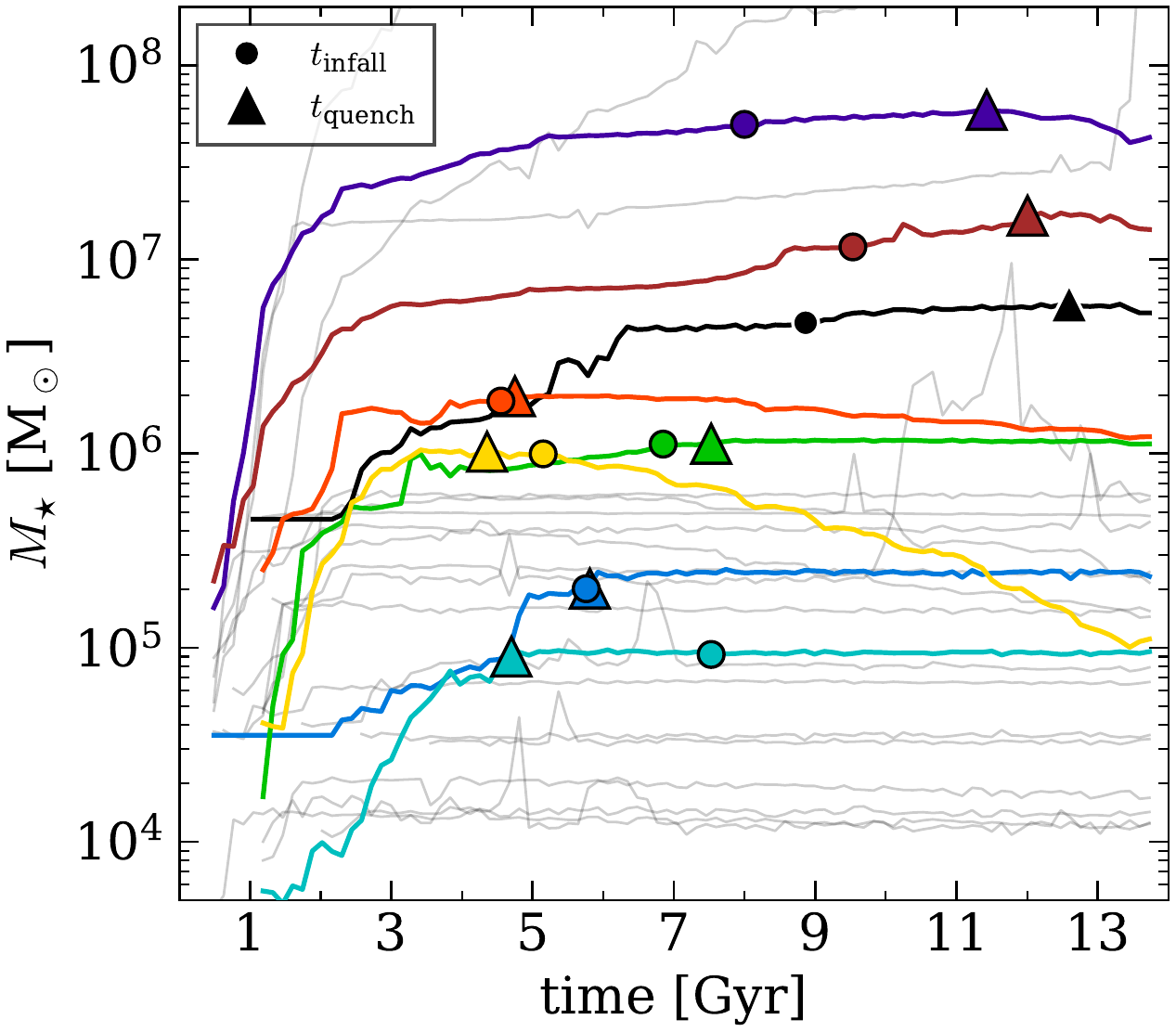}
    \caption{(\textit{Top}) Dark matter and (\textit{bottom}) stellar mass of all satellites versus time, with environmentally quenched satellites highlighted in color. All such satellites experience significant tidal stripping of their DM halos after infall, with anywhere from $82 - 99.9$ per cent of the DM mass being lost by $z=0$. Stellar masses shown here are from \rockstar, and do not necessarily reflect all stellar mass loss due to stripping as streams are not detected and removed. However, stripping of the stellar component can still be seen in objects 82233 \& 66354 (yellow and orange, respectively).}
    \label{fig:m200_mstar_time}
\end{figure}

To investigate the tidal stripping of simulated satellites around LMC-mass hosts, the top panel of Figure \ref{fig:m200_mstar_time} shows the dark matter mass of satellites as a function of time, with the previously described population of environmentally quenched satellites (EQSs) highlighted. We find that the majority of satellites experience tidal stripping of their dark matter halos, beginning at or near their infall times onto the host, with EQSs generally experiencing the largest decreases in halo mass, losing $82 - 99.9$ per cent of the peak halo mass ever obtained.

In one case (halo ID 82233, yellow), the satellite appears to have its halo mass reduced by a factor of $\sim5\times$ prior to quenching, and by $\sim10\times$ prior to first infall. This is the galaxy shown on the bottom panel of Figure \ref{fig:gas_stripping}. It is clear from our previous analysis that ram-pressure plays an important role in its quenching, but here we demonstrate that it is also subjected to severe tidal stripping. This object also experiences the highest magnitude of halo mass loss by $z=0$ due to its short orbital period, early infall time, and apparent interaction with other satellites prior to infall, as seen in Figure \ref{fig:orbits}. 

The bottom panel of Figure \ref{fig:m200_mstar_time} shows the evolution of stellar mass of all satellites. Most satellite galaxies do not experience significant stripping of their stellar components as they are deeply segregated in the inner regions of their dark matter subhalos \citep{penarrubia2008}, but we do find a handful of objects that appear to have had various degrees of stellar mass loss due to stripping. Significant halo mass loss is not necessarily a guaranteed indicator of stellar stripping, but the two halos which lost the highest fraction of halo mass also lost the highest fraction of stellar mass (82233 \& 66354). This makes sense as the DM component is far more extended than the stellar component, and would therefore be first to be stripped away when tidal forces begin to take hold.

\begin{figure}
    \centering
    \includegraphics[width=0.23\textwidth]{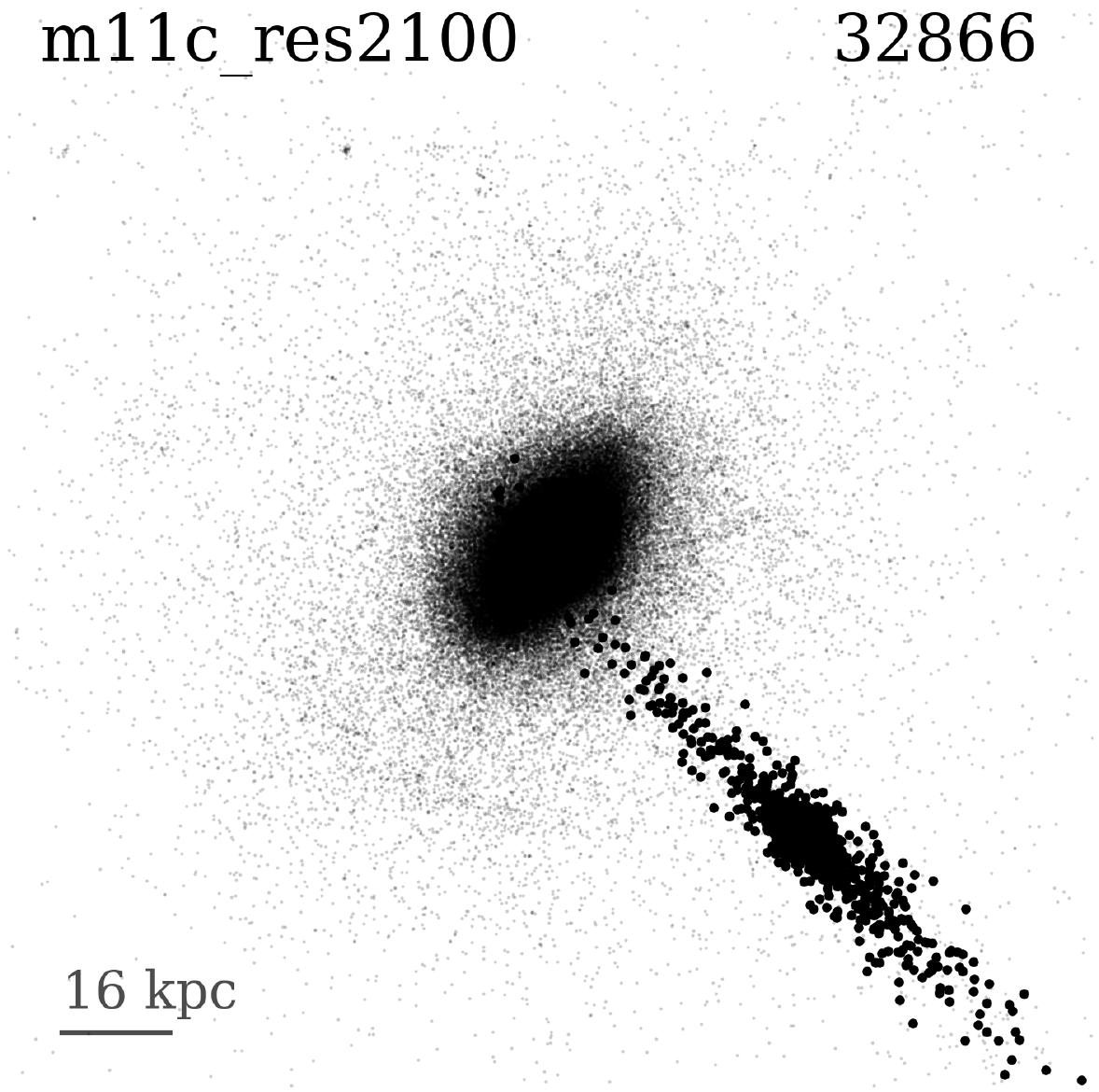}
    \includegraphics[width=0.23\textwidth]{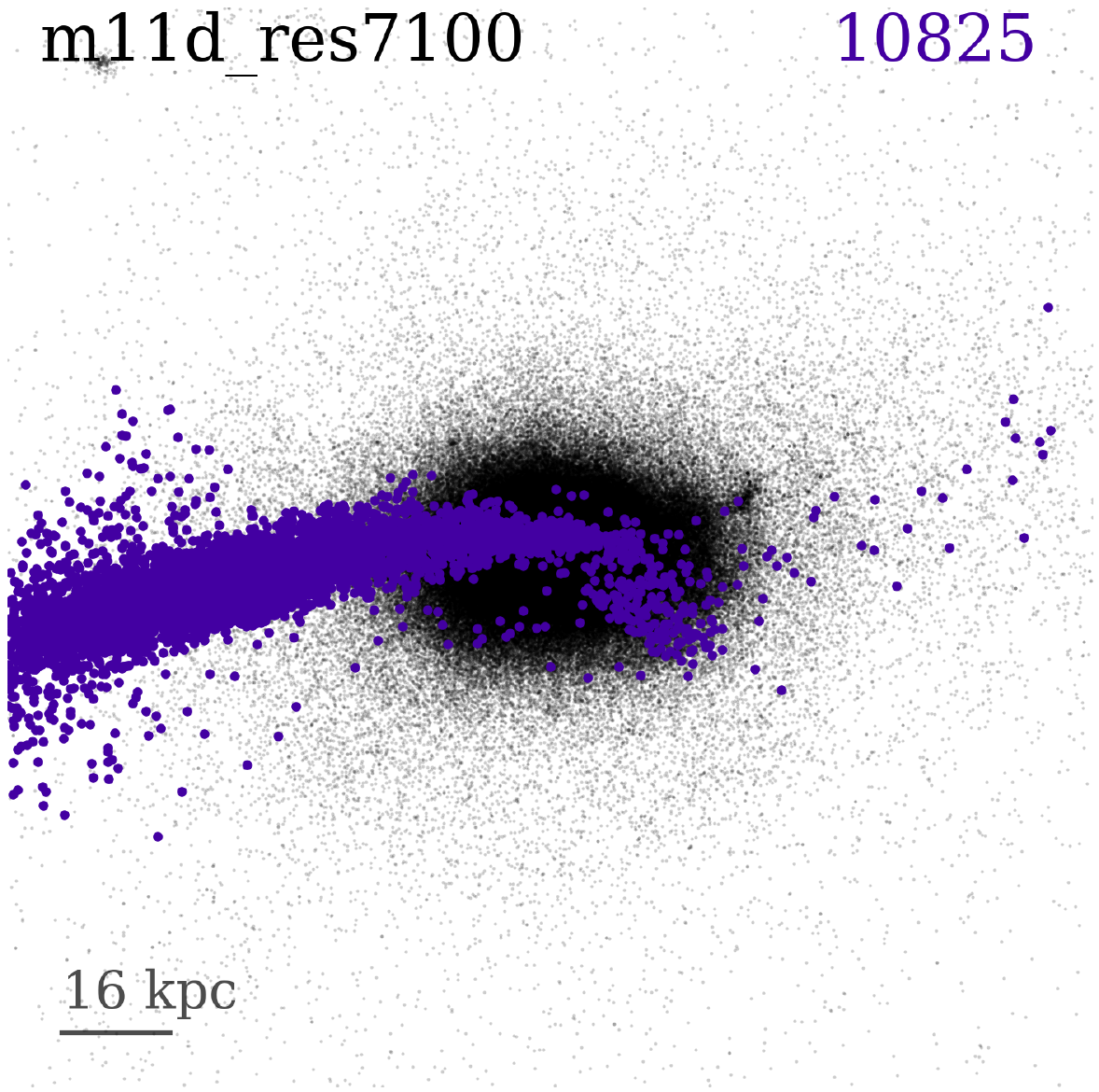}\\
    \includegraphics[width=0.23\textwidth]{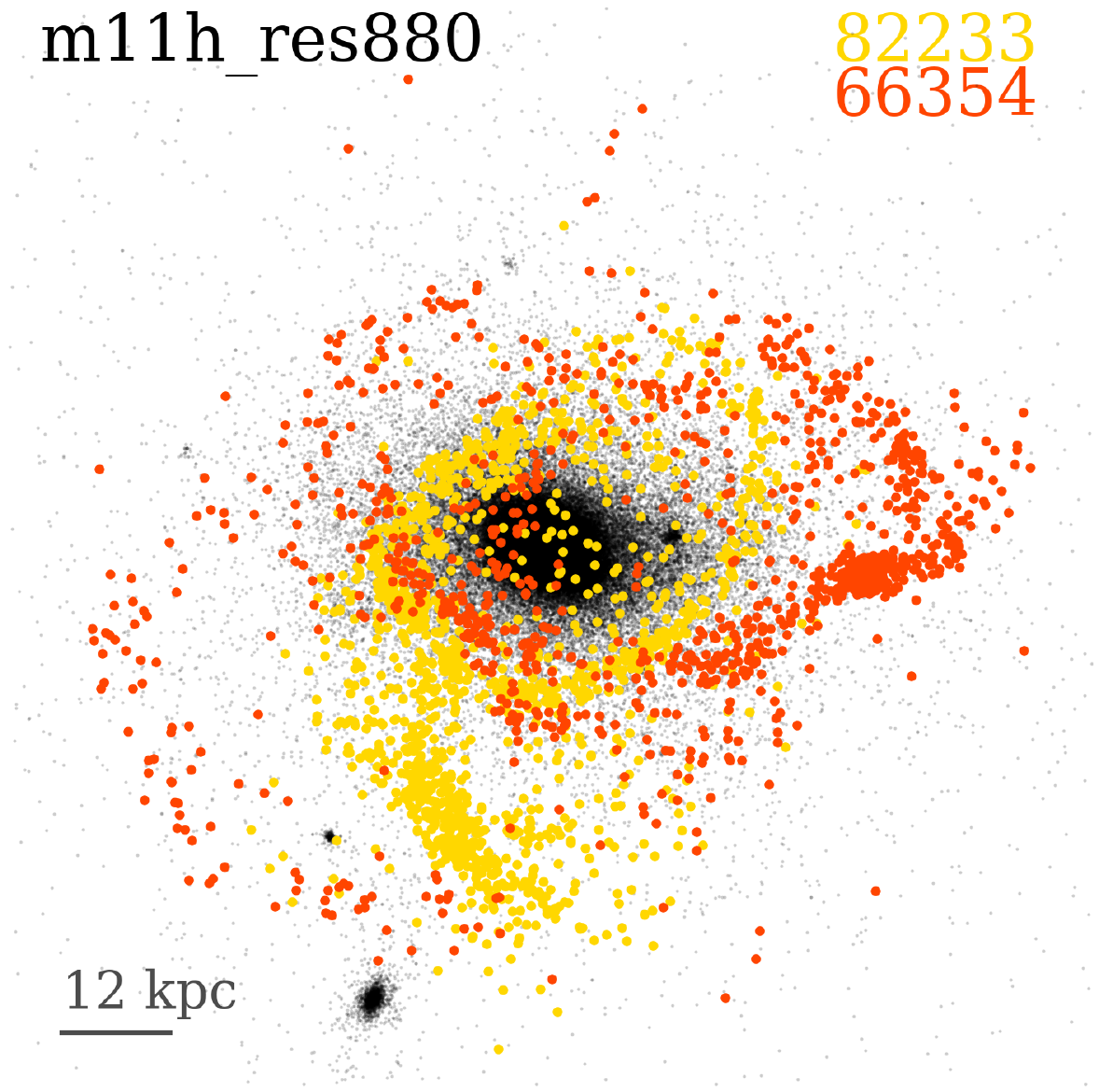}
    \includegraphics[width=0.23\textwidth]{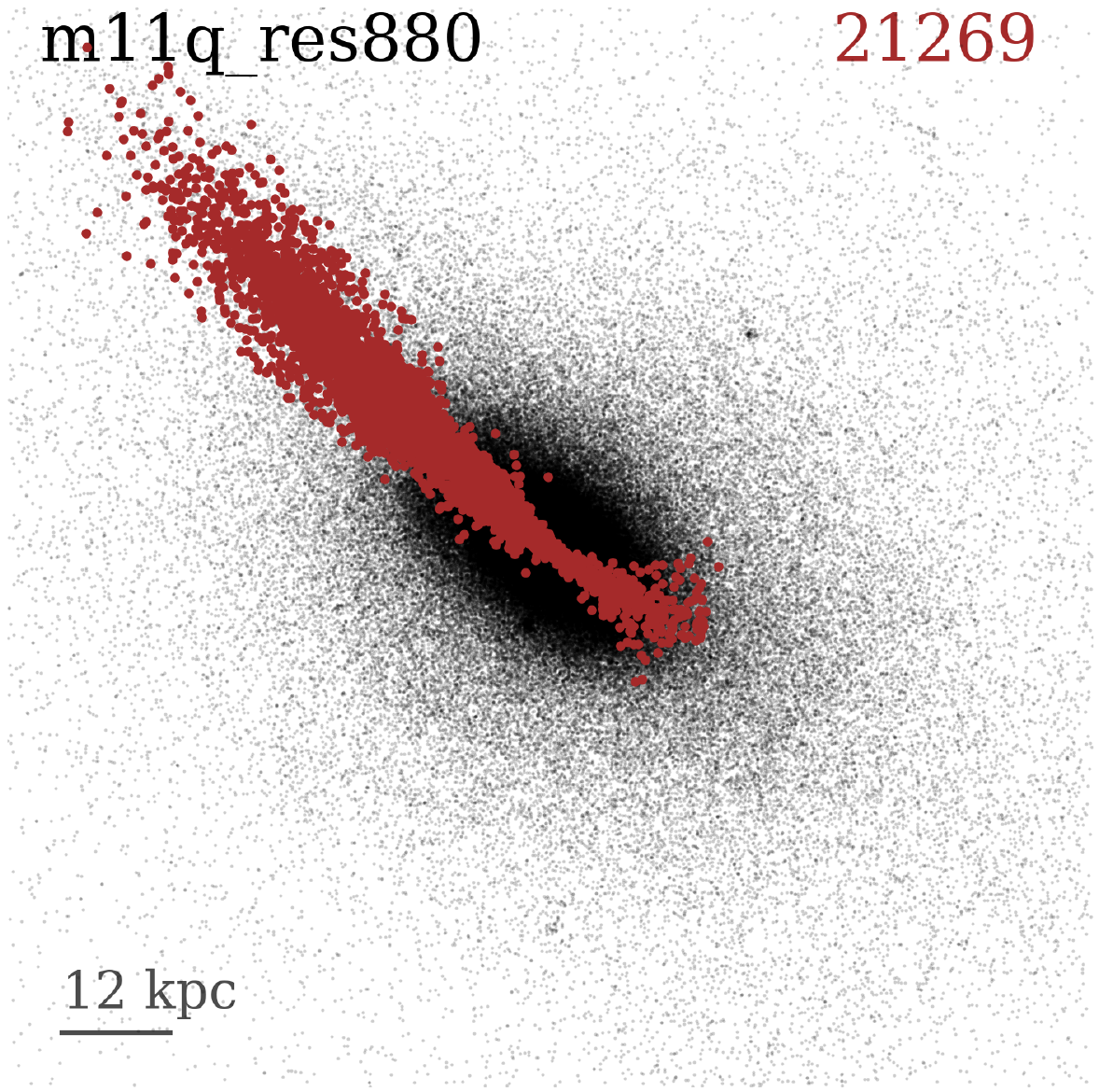}
    \caption{Tidal streams at $z=0$ originating from stripped satellites around LMC-mass hosts. Small, black dots are star particles belonging to the host galaxy at $z=0$, while thicker dots are the present-day locations of star particles belonging to satellites of the corresponding color at their first infall times. We find coherent stellar streams in all hosts with EQSs (Environmentally Quenched Satellites), and none in hosts without EQSs. It is unlikely that the quenching is a direct result of this tidal stripping, but these could be correlated as a result each effect's individual dependence on satellite mass (coherent tidal streams require sufficiently many stars to strip - as well as late(r) infall times - while environmental quenching requires sufficiently high mass as to not by quenched by reionization heating).}
    \label{fig:streams}
\end{figure}

Since gravitational interactions with satellites are known to be a source of stellar tidal streams in MW-mass galaxies, we plot the $z=0$ locations of star particles that were assigned to any satellite galaxies at their infall times in Figure \ref{fig:streams}. Streams were then identified by examining the evolution of the spatial distribution of such star particles. Streams became apparent when star particles were pulled from their original locations within satellites as they made close approaches to the host, forming extended stellar structures. We find four hosts with tidal streams originating from five satellites.

We find no tidal features that arose from any satellite galaxy that was not environmentally quenched, be it a low-mass early quenched galaxy, or a high mass star-forming galaxy. This may be a result of mass selection: low-mass satellites, while occurring frequently and infalling early, do not possess a large population of stars that can be striped with a well-resolved stream in our runs, meanwhile, high mass satellites, while having an abundant stellar component, are less common and infall late, thus not having sufficient time to interact with the host. Tidal stripping is not the dominant factor in quenching these galaxies (see the location of triangle symbols in Fig.~\ref{fig:m200_mstar_time} mostly not correlated to stripping events). It is simply that satellites that were quenched due to ram-pressure stripping seem to be also those experiencing significant tidal stripping not only of their dark matter but also of their stars. 

The streams depicted in Figure \ref{fig:streams} are highly extended, containing stars located within $\sim$1-4 kpc of the primary host galaxy out to $\sim$80 kpc. All stripped stars are on highly radial orbits, in agreement with previous theoretical predictions \citep{Abadi2006}. The amount of stellar mass contained in the streams ranges from $10^{6-7}$ \msun, with a median value of $2\times10^6$ \msun. The streams around our simulated LMC-mass galaxies are quite substantial, and may be observable around dwarf centrals through deep photometry.

We find that tidal structures result from the highest mass environmentally quenched satellites, having a stellar mass range of $M_{\star,\text{max}} = 10^{6\sim 7}$ \msun. Most satellites in this mass range though are star-forming and late-infallers. The tidal structures from later-infalling satellites (such as those around \texttt{m11c}, \texttt{m11d}, and \texttt{m11q}) are morphologically distinct from those formed by early-infalling satellites, as they have not experienced enough dynamical times to become kinematically mixed. The streams around \texttt{m11h} originate from satellites that fell in around 9 Gyr ago, and have undergone many pericenters as seen on Figure \ref{fig:orbits}. This results in streams that are more diffuse, though still retaining clear spatial cohesion along the orbital path.

We note that we only investigate stellar streams that form directly as the result of tidal forces that strip stars from satellite galaxies as they orbit the central. Extended stellar structures also exist in the form of in-situ stellar streams and stellar halos, which have been investigated in the FIRE simulations \citep[e.g.][]{Yu2020}. \citet{elbadry2016} showed that galaxies with $M_\star = 10^{6.3-10.7}$ \msun~experience radial migration of stars on both short and long time-scales due to star-forming clouds that are driven to high radial velocities from bursty feedback, as well as due to energy transfer from the fluctuation of the galactic potential. This migration can result in stars located $\gtrsim 10$ kpc from the radial position of their formation, contributing to wide variations in half light radius over time. They note that the stellar mass range $M_\star = 10^{7-9.6}$ \msun~is optimal for maximizing the physical effects that cause stellar migration, suggesting that LMC-mass centrals may have a significant in-situ stellar halo as well. 

Recently, \citet{Panithanpaisal2021} investigated the formation of stellar streams around MW-mass galaxies in the FIRE-2 simulations. They find that present-day satellites are good proxies for the progenitors of stellar streams. They further show that low mass ($M_\star < 2.25\times10^6$ \msun) stream progenitors are likely to have their star formation quenched prior to infall, while progenitors above that stellar mass threshold are quenched by the host environment. This is consistent with our analysis of EQSs, though we find that low-mass stream progenitors may be environmentally quenched as well. 

While tidal features arising from satellite interactions have been observed around dwarf galaxies \citep[][]{MartinezDelgado2012,Carlin2019}, their frequency in as of yet unknown. The presence of resolved stellar streams in configuration space around four of our six LMC-mass hosts is an encouraging sign that satellite-host interactions may result in observable tidal structures in a substantial fraction of LMC-mass dwarf galaxies in the field. 

\section{Summary \& Conclusions}
We investigate various properties of the satellite population of six LMC-mass hosts in the FIRE simulations. By comparing their star formation histories (SFHs) to those of other centrals of similar stellar mass in Figure \ref{fig:SFH}, we find that LMC satellites have more diverse SFHs and quenching times than central galaxies, which are strongly bimodal -- either forming all their stars before $t=4$ Gyr, or continuing active star formation at $z=0$. We further compare to simulated satellites of Local Group pairs from \citet{GK2019SFH} (also in the FIRE simulations), and find that satellites of LMC-mass hosts have similar SFHs to LG-satellites at fixed mass. LMC satellites retain the general mass-dependence of quenching times: low mass satellites ($M_\star < 10^6$ \msun) quench early, while high mass satellites ($M_\star > 10^7$ \msun) quench late or continue forming stars, as shown in Figure \ref{fig:mstar_fquench}. Intermediate mass satellites have the greatest diversity of quenching times (Figure \ref{fig:mstar_tau90}).

We identified 8 environmentally quenched satellites, selected as having intermediate quenching times (\tquench $=4-13$ Gyr) and located within twice the virial radius of the host at the time of quenching. By examining their orbits and quenching time-scales (Figure \ref{fig:orbits}), we identify two subtypes: higher mass, late infalling satellites that quench after first pericenter; and lower mass, early infalling satellites that quench near the host virial radius. It is unclear whether these subtypes are distinct populations, or if quenching time-scale and infall time are continuous functions of stellar mass. Encouragingly, early data from the LBT-SONG survey also hints at environmental quenching occurring in satellite dwarfs of two observed LMC-like hosts, NGC 628 \citep{Davis2021} and NGC 4214 \citep{Garling2020}. 

All our simulated galaxies experience a stark drop in their gas content after quenching (Figure \ref{fig:mgas}), indicating hydrodynamic rather than gravitational effects. We find that the LMC-mass hosts contain hot, richly structured gaseous halos, with $3\sim6\times10^9$ \msun~of gas in their hot (T $=10^{4.5-6}$K), diffuse ($\rho=10^{-6}$ -- $10^{-4}$cm$^{-3}$) coronas, as shown in Figure \ref{fig:cgm_panel}. We further demonstrate that this rich environment is able to strip gas from satellites via ram-pressure, halting their star formation. This process can be made more efficient through internal burst of feedback within the satellite, moving its gas to a higher energy state and expediting the effects of ram-pressure. Case studies of two satellites that illustrate quenching due to SF-aided ram-pressure stripping versus pure ram-pressure are shown in Figure \ref{fig:gas_stripping}

By examining the evolution of the dark and stellar mass components of satellites, we find that all 8 environmentally quenched satellites have lost $82 - 99.9$ per cent of their peak DM mass via tidal stripping, with other satellites undergoing varying amounts of DM loss, some losing almost none due to their late infall times, as shown in Figure \ref{fig:m200_mstar_time}. Stellar mass loss greater than $\sim$10 per cent due to tidal stripping is rare, happening in only 2 satellites in our sample. 

We investigate vestigial structures of host-satellite interaction by identifying the $z=0$ location of stars that were assigned to satellites at their infall times, and find extended stellar streams around 4 of 6 LMC-mass hosts, seen in Figure \ref{fig:streams}. All originated from environmentally quenched satellites. Three formed from $M_\star = 10^{6.5-7.5}$\msun~satellites infalling within the last $\sim$2 Gyr, while the two streams around the fourth host originated from (pre-infall) $M_\star \approx 10^6$ \msun~satellites with infall time $\sim$8 Gyr ago, around $z\sim1$.

Our findings have strong implications for current and upcoming observational missions targeting LMC analogs in the field. We suggest that such objects may be host to 1-4 intermediate mass ($M_\star = 10^{5-7}$ \msun) satellites which are likely to be environmentally quenched at intermediate -- late times (\tquench$=4-13$ Gyr), depending on mass. This satellite population would be present along side a potential bright, star-forming satellite, as well as $\gtrsim$3 ancient ultrafaint satellites with $10^{4} \leq M_* \leq 10^5$ \msun, though not all of our runs resolve this scale. LMC-mass galaxies in the field can additionally host tidal streams due to past interactions with their satellites.

\section*{Acknowledgements}

LVS acknowledges support from the NASA ATP 80NSSC20K0566 and NSF CAREER 1945310 grants. AW and JS received support from NASA through ATP grants 80NSSC18K1097 and 80NSSC20K0513; HST grants GO-14734, AR-15057, AR-15809, and GO-15902 from STScI; a Scialog Award from the Heising-Simons Foundation; and a Hellman Fellowship. MBK acknowledges support from NSF CAREER award AST-1752913, NSF grant AST-1910346, NASA grant NNX17AG29G, and HST-AR-15006, HST-AR-15809, HST-GO-15658, HST-GO-15901, HST-GO-15902, HST-AR-16159, and HST-GO-16226 from the Space Telescope Science Institute, which is operated by AURA, Inc., under NASA contract NAS5-26555. JSB was supported by NSF grants AST-1910346. We ran simulations using: XSEDE, supported by NSF grant ACI-1548562; Blue Waters, supported by the NSF; Pleiades, via the NASA HEC program through the NAS Division at Ames Research Center.

\section*{Data Availability}
The data supporting the plots within this article are available on reasonable request to the corresponding author. A public version of the GIZMO code is available at \url{http://www.tapir.caltech.edu/~phopkins/Site/GIZMO.html}. Additional data including simulation snapshots, initial conditions, and derived data products are available at \url{https://fire.northwestern.edu/data/}.



\bibliographystyle{mnras}
\bibliography{LMC_environments} 






\bsp	
\label{lastpage}
\end{document}